\documentclass{article}

\usepackage{fullpage}
\usepackage{authblk}
\usepackage{hyperref}
\usepackage{setspace} 

\usepackage{cite}
\usepackage{amsthm}
\usepackage{amsmath}
\usepackage{amssymb}
\usepackage{caption}

\usepackage{microtype}
\usepackage{libertinus}

\usepackage{subfig}
\usepackage{framed}
\usepackage{multirow}
\usepackage{color}
\usepackage{wrapfig}

\usepackage{tikz}
\usetikzlibrary{cd}

\usepackage[inline]{enumitem}
\setlist[enumerate,1]{label={(\arabic*)}}

\newtheorem{theorem}{Theorem}
\newtheorem{definition}{Definition}
\newtheorem{lemma}{Lemma}
\newtheorem{claim}{Claim}

\DeclareMathOperator*{\Exp}{\mathbb{E}}

\newcommand{\id}{\operatorname{id}}
\newcommand{\tab}{\mathrm{tab}}
\newcommand{\Tab}{\mathrm{Tab}}
\newcommand{\man}{\act}
\newcommand{\cvr}{\mathrm{cvr}}
\newcommand{\cvrtransform}{\mathcal{T}}
\newcommand{\act}{\mathrm{act}}

\newcommand{\size}{\mathsf{S}}
\newcommand{\winner}{\mathsf{W}}

\newcommand{\loser}{\mathsf{L}}
\newcommand{\disc}{\mathsf{D}}
\newcommand{\obsdisc}{\mathsf{D}^{\mathcal{A}}}
\newcommand{\obsdiscbatch}{\mathsf{D}^{\mathcal{A}}_{\batchnum}}
\newcommand{\iter}{\mathsf{iter}}
\newcommand{\ballot}{\mathbf{b}}
\newcommand{\batchnum}{\beta}
\newcommand{\groupnum}{\mathsf{g}}
\newcommand{\numgroups}{\nu}
\newcommand{\CVR}{\mathtt{CVR}}
\newcommand{\consistent}{\textsc{Consistent}}
\newcommand{\inconclusive}{\textsc{Inconclusive}}
\newcommand{\RLA}{\mathsf{RLA}}
\newcommand{\Stop}{\mathsf{Stop}}
\newcommand{\Stopmu}{\mathsf{Stop}_{\mu}}
\newcommand{\Stopdelta}{\mathsf{Stop}_{\delta}}
\newcommand{\FirstStop}{\tau_{\Stop}}
\newcommand{\Risk}{\mathsf{Risk}}
\newcommand{\Riskdelta}{\mathsf{Risk}_{\delta}}
\newcommand{\risk}{\alpha}
\newcommand{\Criterion}{\mathsf{R}}
\newcommand{\Criterionmu}{\mathsf{R}_{\mu}}
\newcommand{\Criteriondelta}{\mathsf{R}_{\delta}}
\newcommand{\overvote}{\mathsf{Over}}
\newcommand{\Auditor}{\mathcal{C}}
\newcommand{\Adversary}{\mathcal{A}}
\newcommand{\N}{\mathbb{N}}
\newcommand{\badrow}{\bot}
\newcommand{\corrdisc}{{\disc}^{\cvr,\match}_{r}}
\newcommand{\match}{\mathsf{OneB}}

\begin{document}

\title{Adaptive Risk-Limiting Comparison Audits}

\author{Benjamin Fuller}
\author{Abigail Harrison}
\author{Alexander Russell}

\affil{
  {\small \texttt{\{benjamin.fuller,abigail.harrison,acr\}@uconn.edu}}\\
  Voting Technology Research Laboratory\\
  University of Connecticut}

\maketitle

\begin{abstract}
  Risk-limiting audits (RLAs) are rigorous statistical procedures
  meant to detect invalid election results. RLAs examine paper ballots
  cast during the election to statistically assess the possibility of
  a disagreement between the winner determined by the ballots and the
  winner reported by tabulation.  The design of an RLA must balance
  risk against efficiency: ``risk'' refers to a bound on the chance
  that the audit fails to detect such a disagreement when one occurs;
  ``efficiency'' refers to the total effort
  to conduct the audit.

  The most efficient approaches---when measured in terms of the number
  of ballots that must be inspected---proceed by ``ballot
  comparison.''  However, ballot comparison requires an (untrusted)
  declaration of the contents of each cast ballot, rather than a
  simple tabulation of vote totals. This ``cast-vote record table'' (CVR) is
  then spot-checked against ballots for consistency. In many practical
  settings, the cost of generating a suitable CVR
  dominates the cost of conducting the audit
  which has prevented widespread adoption of
  these sample-efficient techniques.

  We
  introduce a new RLA procedure: an ``adaptive ballot comparison'' audit.
  In this audit, a global CVR is never produced; instead, a three-stage procedure is iterated: 1)
  a batch is selected, 2) a CVR is produced
  for that batch, and 3) a ballot within the batch is sampled, inspected by auditors, and compared with the
  CVR. We prove that such an audit can achieve risk commensurate with
  standard comparison audits while generating a fraction of the CVR.
  We present three main
  contributions:
  \begin{enumerate*}
  \item a formal adversarial model for RLAs;
  \item definition and analysis of an adaptive audit procedure with
    rigorous risk limits and an associated correctness analysis
    accounting for the incidental errors arising in typical audits;
    and
  \item an analysis of efficiency.
  \end{enumerate*}
  
\end{abstract}

\section{Introduction}\label{sec:intro}
\noindent
We consider the task of conducting a risk-limiting audit of a conventional election based on paper ballots. This framework calls for the election to be organized in three stages:
\begin{description}
\item[Ballot casting:] Voters mark paper ballots with their
  preferences, producing a \emph{voter-verified paper
    trail}~\cite{skall2005voluntary,national2018securing}.
\item[Tabulation:] Ballots are
  tabulated and aggregated by
  (untrusted) tabulators  forming
  a tabulated outcome.
\item[Storage:] Ballots are stored in preparation for audits.
\end{description}
The tabulation and storage phases must ensure
``ballot invariance'': no ballots may be destroyed, introduced or modified.
Many countries across the world and
municipalities across the United
States carry out elections modeled on this ideal.

Risk-limiting audits (RLAs) are techniques for testing the veracity of
the tabulation step~\cite{lindeman2012gentle}. Assuming ballot
invariance, RLAs explicitly bound the probability that a disagreement
between the tabulated winner and the
winner determined by the paper trail is undetected by the audit. RLAs
must be transparent: it must be possible for an external observer to
verify that the audit was conducted properly. While a 
variety of specific methods have been proposed, the basic landscape is
dominated by two approaches (see the discussion in
\cite{lindeman2012gentle,RLAWorkbook,rla-working-group,bernhard2021risk,stark2009auditing,stark2010super,higgins2011sharper,lindeman2012bravo,ottoboni2018risk,ottoboni2019bernoulli,stark2020sets,waudby2021rilacs,blom2021assertion} and Section~\ref{ssec:rel work}):
\begin{enumerate*}[label=\alph*)]  \item  ``polling'' randomly sampled ballots to
directly estimate margins, and \item   ``comparing'' randomly sampled
ballots (or groups of ballots) against a \emph{cast-vote record table}. We discuss this approach in detail below.
\end{enumerate*}

As mentioned above, the aim of the audit is to detect circumstances where the tabulated
winner of the election is not, in fact, the winner as determined by
the paper trail.
The paper trail itself---typically consisting of paper ballots marked
directly by voters---is assumed to have an unambiguous interpretation
that serves as the ground truth for the audit.\footnote{In practice, audits may have to contend with disagreements among human interpretations of the paper trail and, in such cases, must provide a mechanism (majority vote, say) for yielding a final interpretation.}
The \emph{risk} of the audit, denoted throughout by $\alpha$, is (an
upper bound on) the probability that the audit incorrectly concludes
an election to be correct when the tabulated and ground truth outcomes
disagree.

\paragraph{Polling.}
A ballot polling audit proceeds by drawing a collection of randomly
sampled ballots; the votes cast on these sampled ballots are then used
to statistically infer the winner of the election. For example, in a
single two-candidate race,
uniform sampling of ballots yields a direct estimate of the
\emph{diluted margin} $\mu$ of the race, equal to the number of votes
cast for the winner minus those for the loser divided by the total
number of ballots cast that contain the race. This estimate achieves risk $\alpha$,  
correctly determining the winner with probability $1-\alpha$,
after sampling
$\Theta(\log(\alpha) / \mu^2)$ ballots.

\paragraph{Comparison.}
Ballot comparison audits, in contrast to polling audits,
require additional metadata about the election: a \emph{cast-vote
  record table} (CVR) that declares the votes cast on each ballot in the
election.
This additional metadata---even though it is not assumed to be
correct by the auditor---yields a dramatic reduction in the number of
ballot examinations necessary for the same risk level: in particular,
only $\Theta(\log(\alpha)/\mu)$ ballots need to be examined to achieve
risk $\alpha$, with $\mu$ as above.\footnote{The use
  of asymptotic notation here is meant to highlight how the efficiency
  of the audit---that is, the number of ballots that must be
  examined---scales with margin.
  Of course, practice demands explicit bounds which have been
  developed by a sizable literature; see~\cite{Stark:Conservative} for
  a survey. We remark that the complexity can also be parameterized in
  terms of the \emph{tabulated diluted margin}, equal to the margin
  defined above with the tabulated vote totals. See~\cite{Stark:2022wt}
  for a detailed discussion.} 

This would appear to establish ballot comparison as the dominant
auditing paradigm as the number of ballots that must be examined
scales more favorably in the margin. However, we are not aware of any
mass-produced \emph{voter-facing} tabulator that produces
ballot-identifying CVRs suitable for a risk-limiting audit. (See the
discussion in Section~\ref{sec:remarks-on-tabulators}.)
For elections with voting facing tabulation, CVRs must then be produced during a second round
of processing by \emph{transitive tabulators} that are specifically
designed to produce CVRs.  (The terminology here is meant to mimic the
language of a ``transitive ballot comparison
audits''~\cite{lindeman2012gentle}.) Unfortunately, this second round
of processing---for reasons we discuss in detail below---tends to
dominate the cost of the ballot comparison audit.\footnote{There are tabulators, such as the ES\&S DS850
  \url{https://www.essvote.com/products/ds850/}, designed for central
  tabulation that produce CVR tables suitable for comparison audits.
  These tabulators directly imprint identifiers on physical ballots in
  order to address the identification problem. Colorado, which uses
  mail-in voting and centrally processes ballots by county, uses such
  tabulators to support 
  ballot comparison audits.}
 
%
 For example, Rhode Island's RLA pilot estimated the setup cost for a
 ballot comparison audit to take roughly six times as long as
 conducting the audit~\cite[Table 2]{RI-Pilot}.\footnote{This assumes
   a 10\% margin and 10\% risk limit with a 75\% chance for the audit to complete.} This was presumably the major
 factor in Rhode Island's adoption of ballot polling (rather than
 ballot comparison) for its RLA of the 2020 presidential
 election~\cite{RI-presidential}. Connecticut's pilot found this ratio
 to be much higher, with CVR generation taking 99\% of the audit
 execution~\cite[Section 6.2]{ctworkinggroup}.\footnote{This analysis
   considers a $2\%$ margin, $5\%$ risk limit, and considers the expected
   number of ballots retrieved. The fraction of time dedicated to CVR
   generation increases as margin increases; one selects fewer
   ballots.}  These pilots used different tabulators
   and different methods for identification---RI imprinted using a high speed scanner, while CT manually applied identifiers. While these figures are from pilots, they indicate that
 CVR generation is an important cost factor in the design and
 implementation of ballot comparison RLAs.

To conclude, ballot comparison audits offer significant
advantages in ballot sample size.  However, in many settings the generation of
CVRs is an expensive, separate step that renders the approach
non-competitive with ballot polling except in circumstances with small
margins. We are not aware of any statewide election procedures in the United States that combine voter-facing tabulators with the efficiency benefits afforded by ballot comparison RLAs.



\subsection{Our results: Adaptive Risk-Limiting Audits}
Typical ballot storage organizes ballots into physical \emph{batches};
in the context of ballot comparison audits, these provide a direct
means for referencing and locating individual ballots. The 
election CVR required for the ballot comparison audit is then
logically composed of a \emph{batch CVR} associated with each
batch.\footnote{For the purposes of this article, the word
``batch'' means a set of ballots that are physically co-located with the standard assumption that
the size of each batch is known with confidence. We also require that each batch has an (untrusted) tabulated
total, which arises naturally when batches are collections of ballots
that were tabulated together (or unions of such collections). 
} To emphasize this distinction, we refer to the full election CVR as a \emph{global CVR}.  In cases where the total number of batches exceeds the number
of ballots sampled during the audit, some batch CVRs will not be
directly examined during the audit procedure. For example, Florida
tabulates by precinct and has over $6000$ precincts~\cite{fl-precinct}.
Even at a 1\% margin, a comparison RLA would only select approximately 20\%
of these precincts for audit (see
Table~\ref{tab:costs}).

\paragraph{Development and analysis of adaptive risk-limiting audits.}
Considering the high cost of CVR generation, we propose an 
``on-the-fly'' procedure for risk-limiting election audits by ballot
comparison. 
 The informal procedure is as follows. (The formal auditor is in Figure~\ref{fig:adaptive-auditor}.)
\begin{enumerate}
\item Ensure that the tabulation is consistent with batch sizes. \label{step:first-check}
\item Repeatedly (or, optionally, in parallel):
\begin{enumerate}
\item Sample a batch with probability proportional to its
  size. Request a CVR to be generated for the
    sampled batch.  (The CVR contains a sequence of rows, each
  containing a ballot identifier and purported votes appearing on the corresponding ballot.)
\item Ensure that the produced batch CVR 
  declares the same total size and votes for the winning and losing candidates as the tabulation of the batch, and declares a 
 unique ballot identifier in each row. \label{step:second-check}
\item Sample a row from the CVR and request a ballot with the identifier appearing in the row. 
\item Compare the retrieved ballot with the votes declared in the CVR row and record their discrepancy. 
\end{enumerate}
\item Compute risk using an appropriate statistical test.
\end{enumerate}
We call this an \textbf{adaptive risk-limiting
  ballot comparison audit} because batch CVRs are created ``on the fly'' and only for batches for which ballot
samples are actually drawn. The audit can additionally incorporate mechanisms to correct consistency failures that might arise in the checks of \ref{step:first-check} and \ref{step:second-check}. The procedure can also benefit from
carrying out CVR generation and sampling for different batches in parallel, known as \emph{audit rounds}.  As such, our techniques are never more costly than a conventional ballot comparison RLA. 

Our main result is a rigorous analysis of the formal procedure which
shows that \emph{with the same number of ballot samples, adaptive
  comparison audits can achieve risk commensurate with standard
  comparison RLAs.}

Adaptive ballot comparison audits can provide significant
  efficiency improvements for RLAs of elections carried out using
  tabulators that do not provide ballot-identifying CVRs (that would directly support comparison RLAs).
Twenty-three of the 50 United States
fall into this category. We use
Connecticut and Florida
as running examples. They differ widely in size: Connecticut is 29th in
population, Florida is 4th.  In addition, Connecticut uses a
transitive tabulator that produces CVRs~\cite{antonyan2013computer}.
Using precinct sizes from the 2020 general election as an
example, for Connecticut, at a $1\%$ margin and 5\% risk limit, $78\%$
of the CVR is generated; for larger margins, as little as $6\%$ of the
CVR is generated.
For Florida, at a $1\%$ margin and 5\% risk limit, only $22\%$ of the CVR is generated; for larger margins, as little as $1\%$ of the CVR is generated. See Table~\ref{tab:costs} for full cost estimates and 
Appendix~\ref{sec:statistical analysis} for justification.

Adaptive RLAs moderate
between the extremes of polling (which is efficient at large margins) and comparison (which is efficient at small margins).  
To explain, the overall time to conduct an adaptive RLA scales with (the
inverse of) margin, while comparison has a large upfront cost to
generate the full CVR and polling requires a sample size that grows
quadratically with (the inverse of) margin.

\begin{table*}[t]
\begin{center}
\begin{tabular}{|r | r | r | r | r| r || r | r | r| r| r|}
        \hline
        &\multicolumn{5}{c||}{$\alpha=5\%$ Risk Limit} & \multicolumn{5}{c|}{$\alpha = 1\%$ Risk Limit}\\\hline
        & \multicolumn{1}{c|}{} & \multicolumn{4}{c||}{Adaptive Comparison} &\multicolumn{1}{c|}{}  & \multicolumn{4}{c|}{Adaptive Comparison}\\
       & &\multicolumn{2}{c|}{CT} & \multicolumn{2}{c||}{FL} &&\multicolumn{2}{c|}{CT} & \multicolumn{2}{c|}{FL}\\
         Margin & Ballots &  Batches& \% CVR   &  Batches& \% CVR   & Ballots  & Batches & \% CVR & Batches & \% CVR
         \\\hline
         1\% & 1532 & 590 & 78\% & 1321 & 22\% & 1886 & 633 & 84\% & 1579 & 26\%
         \\\hline
         2\% & 548 & 331 & 44\% & 515 & 8\% & 725 & 401 & 53\% & 672 & 11\%
         \\\hline
         3\% & 366 & 244 & 32\% & 350 & 6\% & 484 & 304 & 40\% & 458 & 8\%
         \\\hline
         4\% & 274 & 192 & 26\% & 264 & 4\% & 363 & 242 & 32\% & 348 & 6\%
         \\\hline
         5\% & 220 & 160 & 21\% & 213 & 3\% & 290 & 202 & 27\% & 279 & 5\%
         \\\hline
         10\% & 110 & 86 & 11\% & 108 & 2\% & 145 & 109 & 14\% & 141 & 2\%
         \\\hline
         15\% & 74 & 59 & 8\% & 73 & 1\% & 97 & 76 & 10\% & 95 & 2\%
         \\\hline
         20\% & 55 & 44 & 6\% & 54 & 1\% & 73 & 57 & 8\% & 72 & 1\%
         \\\hline
\end{tabular}
\end{center}
\caption{Fraction of CVR generated
  using the Adaptive RLA method for different states, margins, and risk limits. The number of ballot samples is computed with \texttt{rlacalc}~\cite{nealcode}. 
  The percentage of CVR generated by the audit is determined by simulation; see further discussion in
  Appendix~\ref{sec:statistical analysis}. }
\label{tab:costs}
\vspace{-.20in}
\end{table*}

In addition to our adaptive ballot comparison methods, we introduce an
\textbf{adaptive group comparison audit} in
Section~\ref{sec:batch
  comparison} that is intended for settings where ballots are grouped
into small groups (e.g., size $50$) that are interpreted together if
selected. In this setting, no order needs to be kept inside of a group
and ballots do not need to be individually identified.

\subsubsection{The analytic challenge} 
The rigorous analysis of an adaptive ballot comparison RLA must contend with new phenomena
that do not arise in the standard setting: in particular, the batch CVRs
relevant for the audit may be \emph{adaptively} determined as a
function of the entire history of the audit.  Previous analyses also make direct use of the global CVR in order to define
the basic probability-theoretic events of interest; of course, in our
setting this global CVR is not even defined. These considerations lead
to several modeling and analytic challenges, which we briefly summarize.

\paragraph{A formal model for RLAs.} The obligation to rigorously handle
such adaptivity motivates us to lay out a formal model for
risk-limiting audits---borrowing from the successful framework of
cryptographic games---that makes explicit the assumptions and
guarantees offered by the audit. Adopting this model, we then prove
the new procedure is risk-limiting.

\paragraph{Completeness and reflecting ``typical'' auditing errors.}
Such modeling must satisfactorily address the issue of
``completeness,'' by which we mean the ability of the audit to survive
the anticipated errors introduced during practical audit proceedings,
such as occasional inconsistencies in human
ballot interpretation and
mismatches in tabulated batch sizes and
CVR-declared sizes.

\paragraph{Adaptive statistical tests.} Finally, this adaptive setting
places new demands on the underlying statistical tests employed by the
audit. Typical
ballot comparison audits consider tests that
consume \emph{discrepancy vectors} which indicate how selected ballots
differ from the corresponding CVR rows~\cite{lindeman2012gentle}.
In contrast to standard RLA procedures, which can be given a simple analytic treatment in
terms of independent and identically distributed random samples (from
a fixed discrepancy vector), we require tests that provide guarantees
for a broader class of dependent random variables that reflect our
adaptive setting. We formulate a specific ``induced sub-martingale''
condition sufficient for our auditing framework. 
As shown in 
Section~\ref{ssec:concrete tests},
many natural statistical tests satisfy the condition including
the Kaplan--Markov test used in the ``super simple'' ballot comparison
method~\cite{stark2009efficient,stark2009auditing,stark2010super,Stark:Conservative}, the open-source RLA software
\textsc{Arlo},\footnote{\url{https://www.voting.works/risk-limiting-audits}.} and our open-source prototype of the adaptive auditor (\href{https://github.com/aeharrison815/Adaptive-RLA-Tools}{Github repository} and \href{https://mybinder.org/v2/gh/aeharrison815/Adaptive-RLA-Tools/HEAD?labpath=adaptive_notebook.ipynb}{Jupyter notebook}). RLA software design is complex~\cite{bernhard2021risk} and our prototype is meant to inform future development.

\subsubsection{Motivating the formal auditing model}
Our model provides explicit, rigorous answers to natural questions
that may be obscured by informal treatments.
For example:
\begin{itemize}
\item Must ballot identifiers be unique as they appear on physical
  ballots and/or as they appear in a CVR? More broadly, must ballot
  identifiers be determined by trusted auditors?
\item What convention should be adopted for treating mismatches in CVR
  batch size and tabulated batch size?
\item What effect can the---possibly adversarial---destruction
  of ballots have on audit risk and efficiency?\footnote{The reader excited to know
    the answers can refer to Section~\ref{sec:checks needed}.}
\end{itemize}
And, finally, the question that originally motivated the model:
\begin{itemize}
\item What effect can adaptive, adversarial selection of CVRs have on
  audit risk?
\end{itemize}
The model itself introduces two
parties, the \emph{Auditor} and
the \emph{Adversary}. Formally, we consider an election to be defined
by a set of physical ballots and a set of tabulation results (which,
of course, need not match the ballots). The Auditor carries out a
specific, fixed auditing procedure of interest; the Adversary, on the
other hand, is responsible for all of the untrusted aspects of the
audit, such as CVR generation and access to ballots. The notion of \emph{risk}, for a
particular auditor of interest, is now a probability upper bound that
is guaranteed to hold for
all possible behaviors of the adversary.

This corresponds to a guarantee of the risk of the audit even under
situations where a powerful malicious party is attempting to deceive
the auditor; of course, the same guarantees hold in the less
adversarial circumstances that typically hold in practice. The model
also provides a precise method for reasoning about
\emph{completeness}, which reflects the behavior of the audit when
interacting with ``honest adversaries with incidental errors'' that
exhibit the behavior one would expect from tabulators, CVRs, and human
ballot handlers. (See Section~\ref{sec:correctness}.)

\paragraph{Remarks on practical relevance and conventional ballot
  comparison audits.}
Adaptive RLAs will improve efficiency in large-scale elections that \begin{enumerate*}
\item   adopt
tabulators that do not generate CVRs, or tabulators that generate CVRs
without ballot identifying information,
\item  maintain the natural
ballot batching determined by tabulation, which is to say that ballots
tabulated together appear in the same batch,  
\item  yield a number
  of batches that exceeds the anticipated number of sampled ballots, and
\item possess a mechanism to produce CVRs with a corresponding means for identifying individual ballots.
\end{enumerate*}
Currently, 23 US states satisfy these conditions accounting for
roughly half of the US population.  

\paragraph{Remarks on ramifications for conventional comparison audits.} Even in the context of a conventional ballot comparison RLA (in which the full CVR is generated, typically by the tabulator itself),  there are two benefits to these techniques:
  \begin{enumerate}
  \item Our proofs show it is safe to selectively release only the portion of the global CVR corresponding to batches containing selected ballots. This improves the privacy of the audit.
  \item Our model directly specializes to the setting of conventional (non-adaptive) comparison RLAs.  Thus, the fact that uniqueness of ballot identifiers is not necessary for RLA risk guarantees applies to traditional comparison audits as well. To the best of our knowledge, this is the first time this question has been considered. 
  \end{enumerate}

\paragraph{Remarks on tabulators, CVRs and ballot marking.} 
\label{sec:remarks-on-tabulators}
Comparison audits
require a reliable means for identifying specific physical ballots in
order to compare against the CVR. There are two natural means for such
ballot identification: \begin{enumerate*} \item the physical location
  of a ballot and \item identifying marks (``serial numbers'')
  directly printed on ballots. \end{enumerate*} Identifying a ballot
by physical location has typically been implemented by referring to
the \emph{position} of the ballot in a named stack or batch. How this issue is addressed depends on the details of the tabulator. \emph{Voter-facing tabulators} are those that support direct interaction with voters, providing sufficient physical security and privacy features in order for voters to cast their ballots at the tabulator.
Typical voter-facing tabulators intentionally avoid maintaining ballot
order to protect voter privacy; thus the batching of ballots generated
directly from such a tabulator is unsatisfactory for
comparison audits. A further difficulty with ballot position---even
with tabulators that do preserve order---is that the ordering is
transient, subject to corruption during handling, and prone to errors
during ballot indexing;
Colorado, which has successfully used ballot order for identification,
has observed a small but significant error rate~\cite{CO-DiscReport}.

Printing identifying marks directly on ballots addresses these
concerns. However, printing identifiers on ballots \emph{prior} to
voters casting their votes
is a privacy concern. A natural alternative is to indelibly
``imprint'' ballots with identifiers during tabulation. Unfortunately,
this complicates tabulator design: it involves additional hardware
which must provide firm guarantees that marking cannot interfere with
cast vote interpretation and, of course, must not leak voter
  identity.\footnote{The \href{https://dvsorder.org/}{DVSOrder
    vulnerability} is a notable example of an implementation that violated this.} As stated above, these tabulators do not preserve order, so even with identifier imprinting, finding a matching ballot would be complex and time-consuming. This may explain why no mainstream
voter-facing tabulators provide this functionality. These
considerations suggest that the efficiency of near-term ballot
comparison audits with voter-facing tabulators will indeed depend
heavily on CVR generation, which is the principle metric we optimize.

Options for (post-tabulation) CVR generation currently fall into two
categories \begin{enumerate*} \item high-speed, centralized tabulators
  that provide imprinting  and \item
  tabulators specifically designed for transitive use that produce
  CVRs corresponding to ballot identifiers applied in a separate
  ballot identification pass.\end{enumerate*}

Finally, while the election security landscape is complicated, there
are reasons to prefer voter-facing tabulation. Elections are secure and trustworthy when voter registration, authentication, ballot delivery,
vote casting, tabulation, and auditing are tightly coupled. In this context, voter-facing
tabulators provide a strong coupling of vote casting and tabulation.



\subsection{Related work} 
\label{ssec:rel work}
Risk-limiting audits, as the term is now understood, were first
articulated in 2008 by Stark~\cite{starkconservative}. Following this, a body of
work laid down the foundations, including key assumptions and
guarantees~\cite{bernhard2021risk, verifiedvotingprinciples, Hall2009,
  lindeman2012gentle}. As indicated earlier, a
variety of specific methods have been explored, often with an eye to
optimize certain practical settings~\cite{lindeman2012gentle,
  stark2010super, lindeman2012bravo, starkconservative, checkoway,
  starkcast}. A significant literature has also developed around various generalizations and refinements, including
\begin{enumerate*}\item supporting various social choice functions~\cite{stark2020sets,blom2021assertion}, \item
managing multiple races across
jurisdictions~\cite{stark2009auditing,stark2010super,higgins2011sharper,ottoboni2018risk,stark2009efficient}, \item explicit $p$-value estimates~\cite{higgins2011sharper,lindeman2012bravo,ottoboni2019bernoulli,waudby2021rilacs,Stark:Conservative,Stark:2022wt,starkconservative,checkoway,banuelos2012limiting}
and \item implementation
issues~\cite{verifiedvotingprinciples,Hall2009,bernhard2021risk}.\end{enumerate*}

\paragraph{Structure of the paper.} After reviewing preliminaries in Section~\ref{sec:preliminaries}, we present the following:
\begin{enumerate*}
\item an adaptive auditor (Section~\ref{sec:auditor}) that defines the
  details of the adaptive audit procedure;
\item a comprehensive model of election auditing
  (Section~\ref{sec:adversary model}) expressive enough to reflect
  adaptive and traditional comparison RLAs, 
  \item a proof that the adaptive RLA procedure is
  risk-limiting for many existing statistical tests
  (Section~\ref{sec:soundness}), 
  \item a completeness analysis
  establishing that the audits have desirable properties in the
  presence of errors encountered in practical audits
  (Section~\ref{sec:correctness}), and
  \item an adaptive group comparison audit 
  (Section~\ref{sec:batch comparison}). 
\end{enumerate*}

\section{Preliminaries}
\label{sec:preliminaries}
\paragraph{The two-candidate single-race setting.} We consider an audit of a single first-past-the-post
race with two candidates denoted $\winner$ and $\loser$.  By our
naming convention, the candidate $\winner$ is reported to have
received more votes.  The general case---with multiple candidates and
races---can be essentially reduced to this simpler case by
conducting audits for each winner-loser pair simultaneously. The
$p$-values for these can be appropriately combined both for candidate
pairs in the same race and across races. Additional approximations can simplify the
accounting; see~\cite{stark2010super}, which proposes several techniques.

\paragraph{Notation.} We provide a quick overview of notation in
Table~\ref{tab:notation}; this is reviewed as we introduce the adversarial model.  
\begin{table}[t]
\centering
\begin{tabular}{l | l | r} 
& Notation & Description\\\hline
\multirow{6}{*}{concepts}& $\size$ & size\\
& $\winner$ &  tabulated winner\\
& $\loser$ &  tabulated loser \\
& $\mu$ & diluted margin\\
& $\alpha$ & risk limit\\
& $\mathbf{b}, \mathbf{B}_\batchnum$ & physical ballot, batch of ballots\\
& $\disc$ & discrepancy\\\hline
\multirow{3}{*}{modifiers} & $\act$& on ballots\\
& $\tab$ & in tabulation results\\
& $\cvr$ & in CVR
\end{tabular}
\vspace{.05in}
\caption{Notation, reviewed in detail in Section~\ref{sec:preliminaries}.}  
\label{tab:notation}
\end{table}
Throughout, we use boldface to refer to ``physical'' objects, such as
individual ballots (typically denoted $\mathbf{b}$) or groups of
ballots (typically $\mathbf{B}_\batchnum$). Variables determined by these
physical objects are typically denoted with a super- or subscript
($X^{\mathbf{b}}$) with the understanding that they can be determined
from the physical object.

We define $\N = \{0, 1, \ldots\}$ to be the natural numbers (including
zero). For a natural number $k$, we define $[k] = \{1, \ldots, k\}$
(and $[0] = \emptyset$). We let $\Sigma = \{-2, -1, 0, 1, 2\}$, a set
that will play a special role in our setting.
In general, for a
finite set $X$, we define $X^*$ to be the set of all finite-length
sequences over $X$; that is,
$X^* = \{ (x_1, \ldots, x_k) \mid k \geq 0, x_i \in X \}$. Note that
this includes a sequence of length $0$ which we denote
$\lambda$. Finally, we define $X^\N$ to be the set of all sequences
$\{ (x_0, x_1, \ldots) \mid x_i \in X\}$.

\subsection{Election Definitions}
We now set down the elementary definitions of elections, manifests,
and CVRs.
Our setting
demands some generalizations and variants of concepts that are standard in the
literature. In particular, we consider tabulations with batch data and a batch-specific notion of CVR.  See Definition~\ref{def:cvr format} and the preceding discussion.

\begin{definition}[Ballot family; ballot conventions]
  \label{def:ballot-family}
  A \emph{ballot family} is a collection of physical \emph{ballots}
  partitioned into disjoint sets denoted
  $\mathbf{B}_1, \ldots, \mathbf{B}_k$. As a matter of notation, the
  ballot family is denoted
  $\mathbf{B} = (\mathbf{B}_1, \ldots, \mathbf{B}_k)$ and the sets are
  referred to as ``batches.'' For the sake of brevity, we use
  $\mathbf{b} \in \mathbf{B}$ as shorthand for
  $\mathbf{b} \in \bigcup \mathbf{B}_\batchnum$ and use $|\mathbf{B}|$
  as shorthand for $\sum |\mathbf{B}_\batchnum|$.  Throughout, we
  reserve the variable $k$ to refer to the number of batches.

  Physical ballots have three properties:
  \begin{enumerate}
  \item There is an immutable interpretation of the votes contained on
    the ballot.  Each $\mathbf{b} \in \mathbf{B}$
    determines a pair $(\winner_{\mathbf{b}}, \loser_{\mathbf{b}})$,
    where each $\winner_{\mathbf{b}}, \loser_{\mathbf{b}}\in \{0,1\}$.
  \item For any $\mathbf{b} \in \mathbf{B}$, one can determine the batch
    to which the ballot belongs. This defines an index
    $\batchnum_{\mathbf{b}} \in [k]$ such that
    $\mathbf{b} \in \mathbf{B}_{\batchnum_{\mathbf{b}}}$.
  \item Each ballot $\mathbf{b} \in \mathbf{B}$ is \emph{labeled} with
    an indelible \emph{identifier} $\id_\mathbf{b} \in
    \{0,1\}^*$. Ballot identifiers are not necessarily unique; if the
    labels are unique, we say that the family is \emph{uniquely
      labeled}.
  \end{enumerate}
\end{definition}

Some RLAs use the ``location'' of the ballot as the identifier (e.g.,
${\id}_{\mathbf{b}} = \textit{413th ballot in batch 6}$); our framework
works perfectly well in this setting. To reflect practical settings
where certain ballots are actually unlabeled, these can be assigned a
distinguished ``unlabeled'' identifier in $\{0,1\}^*$.
  
\begin{definition}[Tabulation; election]
  \label{def:election}
  Let $\mathbf{B} = (\mathbf{B}_1, \ldots, \mathbf{B}_k)$ be a ballot
  family.  A \emph{tabulation} $T = (T_1,\ldots,T_k)$ for $\mathbf{B}$
  is a sequence where each $T_\batchnum$ is a triple
  $T_\batchnum = (\size^\tab_\batchnum;
  \winner_\batchnum^\tab,\loser_\batchnum^\tab)$ of natural numbers.
  $\size_\batchnum^\tab$ is the number of ballots declared by the
  tabulation in batch $\batchnum$, $\winner_\batchnum^\tab$ is the
  number of votes for the declared winner, and
  $\loser_\batchnum^\tab$ is the number of votes for the declared
  loser.  For a tabulation $T$, the \emph{tabulated totals} are
  \[
    \winner^{\tab} = \sum_\batchnum \winner_\batchnum^{\tab}\qquad \text{and}\qquad
    \loser^{\tab} = \sum_\batchnum \loser_\batchnum^{\tab}\,
  \]
  with the convention that $\winner^{\tab} > \loser^{\tab}$.
  
  An \emph{election} $E$ is a pair $E = (\mathbf{B}, T)$ where
  $\mathbf{B}$
  is a ballot
  family and $T = (T_1,..., T_k) $ is a \emph{tabulation} for $\mathbf{B}$. 
\end{definition}

We do not treat elections that declare a tie between $\winner$ and
$\loser$, with the assumption that this would result in a runoff
or a full hand-count audit.

\smallskip
\noindent
\emph{Notational warning.} The candidate $\winner$ is the \emph{declared
  winner} of the election (according to the tabulation). The
tabulation may not, of course, accurately reflect the votes recorded
on the ballots. The primary circumstance of interest arises when
$\winner$ is not the true winner of the election.

\begin{definition}[Actual vote totals; ballot manifests]
  Let $E = (\mathbf{B}, T)$ be an election.
Let
\[
((\size_1^\act; \winner_1^\act, \loser_1^\act), \ldots, (\size_k^\act; \winner_k^\act, \loser_k^\act))
\]
   denote the actual totals, where $\size_\batchnum^\act = |\mathbf{B}_\batchnum|$ is the
  actual size of batch $\batchnum$ and
  \[
    \winner^\act_\batchnum = \sum_{\mathbf{b} \in \mathbf{B}_\batchnum} \winner_{\mathbf{b}} \qquad\text{and}\qquad \loser_\batchnum^\act = \sum_{\mathbf{b} \in \mathbf{B}_\batchnum} \loser_{\mathbf{b}}
  \]
  are the total number of actual votes received by candidate $\winner$
  and candidate $\loser$ in batch $\batchnum$.  The actual totals are
  \[
    \winner^{\act} = \sum_\batchnum \winner^{\act}_\batchnum \qquad\text{and}\qquad \loser^{\act} = \sum_\batchnum \loser^{\act}_\batchnum\,.
  \]
  The \emph{ballot manifest} of $E$ is the tuple
  $\size_E^\act = (\size_1^\act, \ldots, \size_k^\act)$.
\end{definition}

\begin{definition}[Diluted margin; valid and invalid elections] The 
  \emph{tabulated diluted margin} of an election $E$ is the quantity
  \[
    \mu^\tab = \frac{\winner^\tab - \loser^\tab}{|\mathbf{B}|}\,.
  \]
  An election $E$ is \emph{invalid} if the tabulated winner is
  incorrect: $\loser^{\act} \geq \winner^{\act}$; otherwise, we say
  that $E$ is \emph{valid}.

The tabulated diluted margin is determined by both the number of physical ballots (as determined by the ballot manifest) and the tabulation; to
  emphasize this, we use the notation $\mu^\tab$.  This is in contrast to the \emph{actual diluted margin}
$
    \mu^\act = | \winner^\act-\loser^\act |/|\mathbf{B}|
  $ which is determined only by the physical ballots.
\end{definition}


\paragraph{Cast-vote records (CVRs).}
\begin{wrapfigure}{r}{33mm}
  \centering
\begin{tabular}{|c|c|c|}
  \hline
  Ident. & $\winner$ & $\loser$\\ \hline\hline
  $\id_1$ & 1 & 0 \\ \hline
  $\badrow_1$ &1 & 0 \\ \hline
  $\id_3$ & 0 & 1\\ \hline
  \vdots &\vdots& \vdots \\\hline
\end{tabular}
\caption{A CVR.}
\end{wrapfigure}
A cast-vote record table (CVR) is an (untrusted) declaration of both the
ballots appearing in a particular physical batch and the votes
appearing on the ballots. Each row of the CVR contains a ballot identifier and two entries in
$\{0,1\}$ indicating whether the purported ballot contains a vote for
$\winner$ or $\loser$. 

In our setting, it is critical that tabulations provide batch-level subtotals which can be compared against the totals declared by adaptively generated CVR tables. Traditional RLAs require only a ``global'' CVR and the global consistency check that it induces the same winners and losers as the tabulation.

\begin{definition}[Cast-Vote Record Table (CVR)] Let $\mathbf{B}$ be a
  ballot family. A
  \emph{Cast-Vote Record Table (CVR)} for batch $\batchnum$ is a
  sequence of triples
  \[
    \cvr = ((\iota_1, \winner_1, \loser_1), \ldots, (\iota_s,
    \winner_s, \loser_s))
  \]
  where each $\iota_r$ is a bitstring in $\{0,1\}^*$ and each
  $\winner_r, \loser_r$ is an element of $\{0,1\}$. We use the following language:
  \begin{enumerate}
  \item The elements
    $\iota_r$ are \emph{identifiers}.
  \item The number $s$ is 
  the \emph{size} of the CVR.
  \item The $r$th row  is a triple
  $\cvr_r = (\iota_r, \winner_r, \loser_r)$.
\item 
The values
  \[
    \size^\cvr_\batchnum = s, \; \winner^\cvr_\batchnum = \sum_{1
      \leq r \leq s} \winner_r^\cvr, \qquad \text{and} \qquad
    \loser_\batchnum^\cvr = \sum_{1
      \leq r \leq s} \loser^\cvr_r\,.
  \]
  These denote the number of ballots declared by the CVR and the
  number of votes declared for the two candidates in the CVR.
\item If the identifiers appearing in the CVR are unique, we say the
  CVR is \emph{uniquely labeled}. If a CVR is uniquely labeled we use $r_\iota$
  to refer to the (unique) row with identifier $\iota$. Looking ahead, in Figure~\ref{fig:transforms} we use the identifiers $\perp_i$ to transform a CVR to one with unique labels; such labels would not appear on CVRs generated by tabulators.
  \end{enumerate}
  \label{def:cvr format}
  Finally, a sequence $(\cvr_1, \ldots \cvr_k)$, where each
  $\cvr_\batchnum$ is a CVR for batch $\batchnum$, is a \emph{global
    CVR}.  
\end{definition}

\noindent

\paragraph{Discrepancy.}
\label{sec:discrepancy}
Discrepancy measures the disagreement between claimed
vote tallies, either from a tabulation or CVR, and vote tallies
determined by actual ballots.
\begin{definition}[Batch and election discrepancy] Let
  $E = (\mathbf{B}, T)$ be an election. The \emph{discrepancy} of a batch $\mathbf{B}_\batchnum$ is
\[
  \disc_\batchnum = (\winner^\tab_\batchnum - \loser^{\tab}_\batchnum) - \sum_{\mathbf{b} \in \mathbf{B}_\batchnum}\left(\winner^\act_\mathbf{b} - \loser^{\act}_\mathbf{b}\right) \,.
\]
The overall discrepancy of an election is 
\[
  \disc = \sum_\batchnum \disc_\batchnum = (\winner^\tab - \loser^{\tab}) - (\winner^\act - \loser^{\act})\,.
\]

\end{definition}
\noindent
For invalid elections $\loser^\act\ge \winner^\act$ and thus $\mu^\act = -(\winner^\act- \loser^\act)/|\mathbf{B}|$. In this case
\begin{align}
\frac{\disc}{|\mathbf{B}|} = \frac{(\winner^\tab - \loser^{\tab}) - (\winner^\act - \loser^{\act})}{|\mathbf{B}|} = \mu^{\tab} + \mu^\act.
\label{eq:disc to margin tab}
\end{align}

\noindent
The discrepancy of a CVR is undefined until it is generated,
which is why the above ``global'' definitions focus on the tabulation.

\begin{definition}
  [CVR Discrepancy]
  \label{def:discrepancy}
  Let $\mathbf{B}$ be a ballot
  family and let
  $
    \cvr =
    ((\iota_1,\winner_1,\loser_1),\ldots,(\iota_s,\winner_s,\loser_s))
  $
  be a CVR for batch $\batchnum$. For a row $r \in [s]$,
  define the \emph{discrepancy} $\disc_r^\cvr$ of the row $r$ to be the value
  \begin{equation}
    \label{eq:row-disc}
    \winner_{r} - \loser_{r}
     + \min \left( \{1\} \cup \bigl\{ -
                     (\winner_\mathbf{b} - \loser_{\mathbf{b}}) \mid \id_{\mathbf{b}} =
                     \iota_r, \mathbf{b} \in \mathbf{B}_\batchnum\bigr\}\right) \,.
  \end{equation}
  The minimum is taken over all ballots for which
  $\id_{\mathbf{b}} = \iota_r$ with the default value of $1$
  (intuitively corresponding to a ``concealed vote'' for the declared
  loser) when no ballot corresponds to the identifier.
\end{definition}

When discrepancy takes a positive value $d$ we refer to it as a
\emph{$d$-vote overstatement}; likewise, when it takes a negative
value $-d$ we refer to it as a \emph{$d$-vote understatement}.  In the
context of a tabulation, then, a $d$-vote overstatement indicates that
the reported difference, $\winner^\tab - \loser^\tab$, is $d$ votes
too large.
Equation~\eqref{eq:row-disc} assigns a notion of discrepancy to a
particular row of a CVR, which always takes a value in the set
$\Sigma = \{-2, -1, 0, 1, 2\}$.  In the case
when an identifier $\iota$ corresponds to a unique ballot
$\ballot$, the discrepancy is the natural difference
\[
  \winner_{r_\iota} - \loser_{r_\iota} - (\winner_{\mathbf{b}} -
  \loser_{\mathbf{b}})\,.
  \]

\section{The Adaptive Auditor}
\label{sec:auditor}
\begin{figure*}
  \centering
     \subfloat[][Architecture of Ballot Comparison Audit with Rounds]{\includegraphics[width=3in]{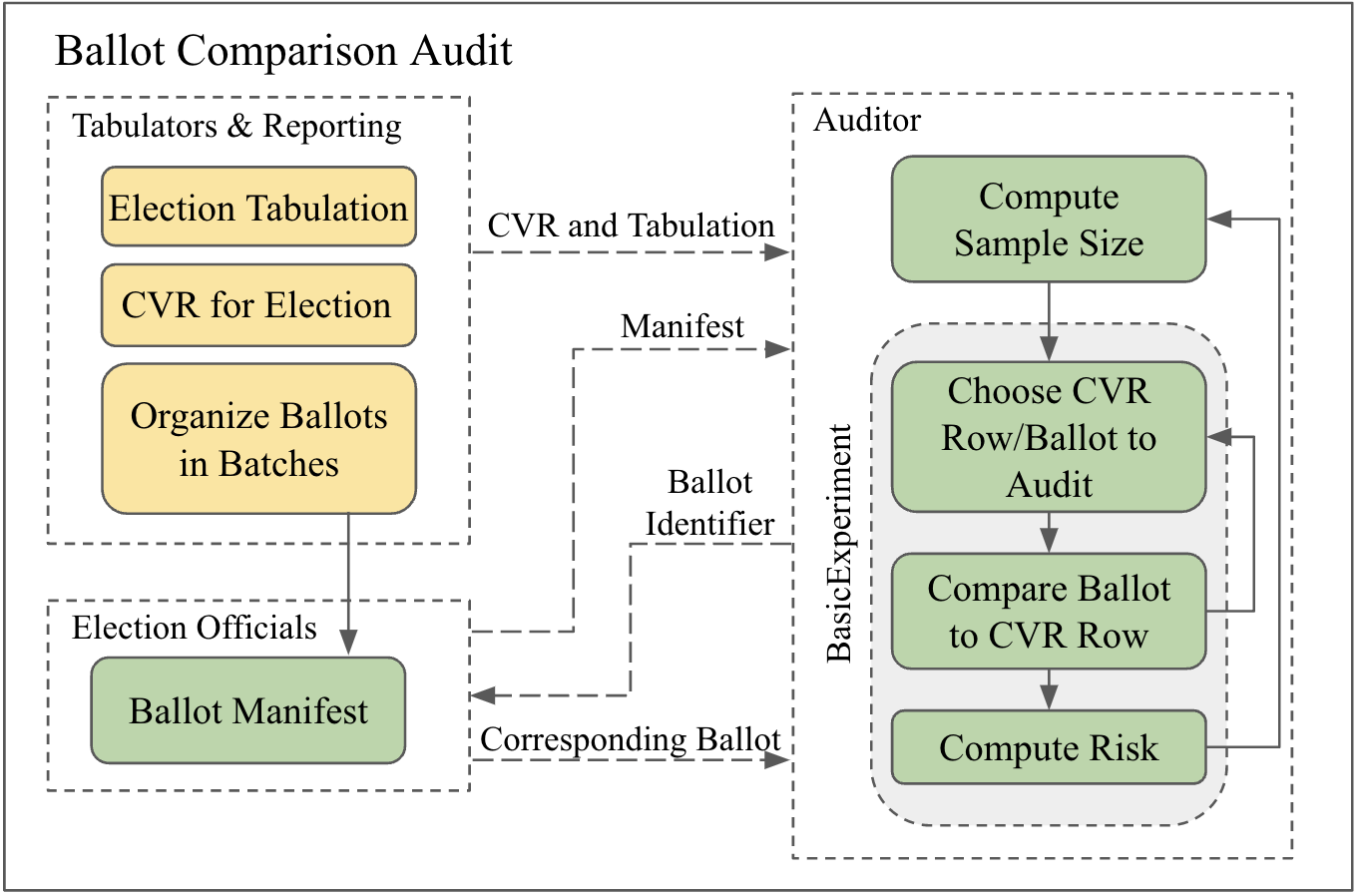}\label{fig:rlaarch}}
     \hspace{.3in}
     \subfloat[][Architecture of Adaptive Ballot Comparison Audit with Rounds]{\includegraphics[width=3in]{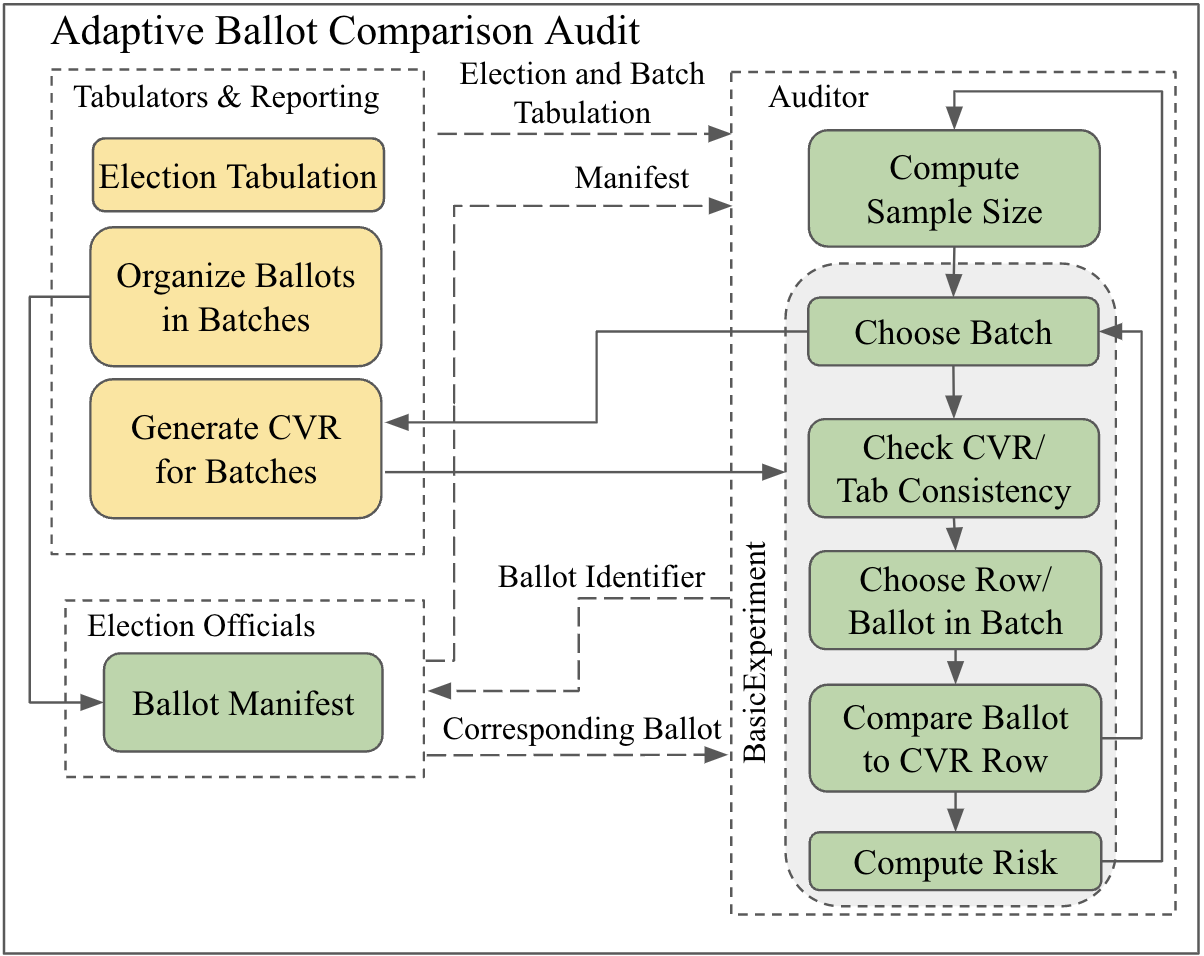}\label{fig:adaptiverlaarch}}

     \caption{Comparison of traditional and adaptive ballot comparison architectures.  Yellow components are performed by untrusted components.  Green components must be trustworthy. The dotted arrows represent information trade, while the solid arrows are procedure steps. The grey procedure is $\mathtt{BasicExperiment}$ can be done in parallel in both traditional and adaptive RLAs. Note that in a traditional audit, the CVR is generated as part of the audit process; in the adaptive setting, the CVR is generated only as the auditor chooses batches. The step of checking CVR and tabulation consistency is also absent from traditional comparison audits as an audit of the CVR is an audit of the tabulation as long as they show the same set of winning/losing candidates.}
     \label{fig:rla architectures}
\end{figure*}

A traditional ballot comparison audit proceeds as follows (illustrated in Figure~\ref{fig:rlaarch}):
\begin{enumerate}
\itemsep0em
\item \label{step:election} An election is carried out, electronic tabulators generate an
  \emph{untrusted} tabulation. 
  \item Election officials store the physical ballots as a ballot family and produce a \emph{trusted} ballot manifest that correctly indicates the
    number of physical ballots in the batch. \label{step:manifest}
\item An \emph{untrusted} CVR is
  generated.
\item The audit repeatedly selects a CVR row and ensures that the
  corresponding physical ballot matches the declaration of votes on
  the CVR row.
  \end{enumerate}
  The audit either generates a risk-controlled declaration that the
  tabulated outcomes are consistent with the ballots or an
  inconclusive result.
%

\paragraph{The adaptive alternative.} As described in the introduction we consider the \emph{adaptive} version of the above (shown in Figure~\ref{fig:adaptiverlaarch}) where CVRs are only generated when needed. This yields the
following family of auditing procedures.
\begin{enumerate}
\item An election is carried out and ballot family created as in steps~\ref{step:election}-\ref{step:manifest}
  above. The tabulation declares a (sub-) tabulation for each batch in the ballot family.

\item The audit consists of multiple instances of the following \textbf{basic
  experiment}, which may be carried out in parallel:
  \begin{enumerate}
  \item A batch is sampled with probability proportional to the number of ballots.
  \item \label{step:CVR} An (untrusted) CVR is generated for the batch.
  \item The CVR is compared against the 
    declared subtotals.
  \item \label{step:compare} An entry in the CVR is drawn uniformly
    and compared with the corresponding ballot.
  \end{enumerate}
\end{enumerate}
As above, the conclusion is either ``consistent'' or
``inconclusive.''
  
Multiple iterations of the basic experiment can be performed in parallel
as in a traditional ballot comparison audit to allow audit workers to
create their portion of the CVR simultaneously.  These are known as audit \emph{rounds} which yield a trade-off between the total number of
examined ballots and the probability of carrying out an additional
round of auditing. The impact of conducting multiple
rounds can be quite high, so parameters are typically chosen to ensure a
single-round audit with high probability. All of this existing
machinery applies identically in our setting.

This section focuses on the audit procedure.  However,  a few
preliminary remarks about modeling are in order. The risk guarantee associated with a standard comparison audit must
hold for all possible CVRs that could be submitted for the election,
even those that might be specifically designed to frustrate the audit
or obscure an invalid election. This motivates our treatment of the
environment in which an auditor operates as \emph{adversarial},
including the CVRs that are produced. We additionally assume an arbitrary 
labeling of ballots.

The informal treatment above already highlights an important
difference between conventional comparison audits and adaptive audits: the
CVR generated and used for comparison by the auditor in steps \ref{step:CVR}--\ref{step:compare}
may depend on the prior history of the audit.  The need to bound risk must hold when the CVRs proposed at
intermediate steps of the audit might depend adversarially on prior
CVRs, row selections, and comparison results. This ability of an
adversary intent on concealing an invalid election appears to be very
powerful: for example, if an adversary has been ``caught'' in a
comparison iteration they may choose to declare subsequent CVRs with a
low discrepancy in order to convince the statistical test that
``everything is OK.''  The above
procedure appears to be
the first RLA involving an adaptive
adversary that engages with the auditor.




We begin by introducing a ``strict'' auditor that enforces size
checks, insisting that the CVR is consistent with tabulation. This
auditor is not necessarily useful in practice, but is a convenient
analytic tool. We then generalize this auditor by defining the notion
of a \emph{CVR transform function} that is applied before the auditor
checks consistency. This extra flexibility makes it easy to construct
and reason about more permissive auditors that are useful in
practice. As we show in Lemma~\ref{lem:permissive okay}, if the original strict auditor
(with the identity CVR transform) is risk-limiting then the resulting
auditor is risk-limiting for \emph{every} CVR transform. This allows us to introduce a transform that always produces ``consistent'' CVRs.

In the next three  subsections, we discuss single-tailed
statistical tests, the auditor, and the intuition for included checks.
We then present the formal game including the definition of risk limit in Section~\ref{sec:adversary model}, show that the auditor is risk-limiting for an
appropriate statistical test in Section~\ref{sec:soundness}, and discuss
completeness in Section~\ref{sec:correctness}.

\subsection{Adaptive single-tailed statistical tests} 
\label{ssec:statistical tests} A standard approach for designing RLAs
is to consider the discrepancy
$\disc_r^\cvr = (\winner_r - \loser_r) - (\winner_{\mathbf{b}} -
\loser_{\mathbf{b}})$ of a uniformly selected row $r$ of a global CVR
in comparison
with a ballot $\mathbf{b}$ corresponding to this entry (as in
Definition~\ref{def:discrepancy}). In light of Equation~\ref{eq:disc
  to margin tab}, if the election is invalid one has that
\[
  \Exp_r[\disc_r^\cvr] \ge \mu^\tab +\mu^\act \ge \mu^\tab\,.
\]
Independently repeating this experiment results in
a sequence of discrepancy observations $\disc_1, \disc_2,...$ taking values in
$ \{-2,-1,0,1,2\}$.  With these random variables, one can formulate an
RLA as a conventional
statistical hypothesis test by adopting
the null hypothesis that the election is invalid;
then one is interested in bounding the probability that the null
hypothesis is rejected when it is true. An RLA is determined by
a single-tailed statistical test for these i.i.d.\ random variables
with the hypothesis that ``$\Exp[\disc_i] \geq \mu^\tab$.''  The test
decides whether to reject this hypothesis based on examination of a
finite-length prefix $\disc_1, \ldots, \disc_\tau$ of the variables
given by a ``stopping time.''  Informally, such a test has \emph{risk}
(Type I error) $\alpha$ if
$\alpha \geq \Pr[\text{hypothesis rejected}]$ when indeed
$\Exp[\disc_i] \geq \mu^\tab$. See \cite[Equation 5]{stark2010super}
for further discussion.

\paragraph{The adaptive setting and the domination inequalities.}
In our setting with an adaptive adversary, we will require statistical
tests with stronger properties. Specifically, as above we consider an
infinite family of random variables $X_1, X_2, \ldots$ taking values
in $\Sigma$ with the weaker \emph{domination} conditions recorded
below.

\begin{definition}[$\delta$-dominating distributions and random variables]
  A sequence of bounded (real-valued) random variables $X_1, \ldots$
  are said to be \emph{$\delta$-dominating} if, for each $t \geq 0$,
  \[
  \Exp[X_t \mid X_1, \ldots, X_{t-1}] \geq \delta\,.
  \]
  We also use this terminology to apply to the distribution
  $\mathcal{D}$ corresponding to the random variables, writing $\delta \unlhd \mathcal{D}$.
\end{definition}

The variables are no longer required to be independent or have
the same distribution; however, they still possess the property that
under any conditioning on the past, each random variable has
expectation bounded below by $\delta$. 

\begin{definition}[Stopping time] Let $\Sigma = \{-2, -1, 0, 1,
  2\}$. A \emph{stopping time} is a function
  $\Stop: \Sigma^* \rightarrow \{0,1\}$ so that for any sequence
  $x_1, x_2, \ldots$ of values in $\Sigma$ there is a finite prefix
  $x_1, \ldots, x_k$ for which $\Stop(x_1, \ldots, x_k) = 1$.
    
  For a sequence of random variables $X_1, \ldots$ taking values in
  $\Sigma$, let $\FirstStop(X_1, \ldots)$ be the random variable given
  by the smallest $t$ for which $\Stop(X_1, \ldots, X_t) = 1$. This
  naturally determines the random variable
  $X_1, \ldots, X_{\FirstStop}$, the prefix of the $X_i$ given by the
  first time $\Stop() = 1$.
  \label{def:stop time}
\end{definition}

With these preliminaries noted, we can define the family of
statistical tests that we show can support adaptive audits.

\begin{definition}[Adaptive Audit Test]\label{def:audit-test}
  An \emph{adaptive audit test}, denoted $T = (\Stop, \Criterion)$, is
  described by two families of functions, $\Stopdelta$ and $\Criteriondelta$. For
  each $-2 \le \delta \le 2$,
  \begin{enumerate}
  \item $\Stopdelta$ is a stopping time, as in Definition~\ref{def:stop time}, and 
  \item $\Criteriondelta:\Sigma^* \rightarrow \{0,1\}$ is
    the \emph{rejection criterion}. 
  \end{enumerate}
  Let $\mathcal{D}$ be a probability distribution on $\Sigma^\N$; for
  such a distribution, define
  $\alpha_{\delta,{\mathcal{D}}} = \Exp[\Criteriondelta(X_1, \ldots,
  X_{\tau})]$ where $X_1, \ldots$ are random variables
  distributed according to $\mathcal{D}$ and $\tau$ is determined by
  $\Stopdelta$.
  Then we define the \emph{risk} of the test to be
  \begin{equation}
  \label{def:audit test}
    \alpha = \sup_{\substack{0 < \delta < 2\\ \delta \unlhd \mathcal{D}}} \alpha_{\delta,{\mathcal{D}}},
  \end{equation}
  where this supremum is taken over all $\delta \in (0,2]$ and over all probability distributions
  $\mathcal{D}$ for which $\delta \unlhd \mathcal{D}$.
  %
\end{definition}

In Section~\ref{sec:soundness} we observe that several
families of statistical tests in common use---including the popular Kaplan-Markov test---are, in fact, adaptive
audit tests.

\subsection{The Adaptive Audit Procedure}
We now present the adaptive auditor (Figure~\ref{fig:adaptive-auditor}).  The design of the audit procedure is motivated by three guiding principles:
\begin{enumerate}
\item Ensure tabulation consistency with the ballot manifest.  (This means the size must match, $\winner^\tab\le \size^\act$, and $\loser^\tab\le \size^\act$).  Such checks ensure that the overall discrepancy is at least the margin for invalid elections.  This principle motivates 
  Steps~\ref{step:correct sizes} and \ref{step:compute margin}.
\item Ensure that duplicate labels appearing on distinct ballots cannot
  increase risk. This follows from (i.) forcing CVR tables to contain no duplicates, (ii.) adopting uniform selection of CVR rows for ballot selection and, (iii.) noting that among the collection of ballots that may be assigned a common identifier, there is a ``pessimal'' ballot that induces the minimum discrepancy.
  See $\mathtt{CheckConsistent}$ and Step~\ref{step:ballot id check}
  of $\mathtt{BasicExperiment}$.
\item Ensure that any produced CVR for a batch has the same number of votes for the winner and loser as the declared tabulation for that batch.  This yields a lower bound on the discrepancy---determined only by the tabulation and the ballots---between any such CVR and the ballots.
  See  the additional checks in $\mathtt{CheckConsistent}$.
\end{enumerate}

This auditor and the related treatment of ballot identifier uniqueness also have direct ramifications for traditional comparison audits; see the discussion in Section~\ref{ssec:conventional rlas} below.

\begin{figure}[th!]
  \begin{framed}
    \underline{Auditor $\Auditor[\cvrtransform;(\Stop, \Criterion)]$ for an
    election $E$}
    \begin{enumerate}[noitemsep]
    \item Receive ballot manifest and tabulation:
            \begin{align*}
              \size_E^\act &= (\size_1^\act, \ldots, \size_k^\act);
     \qquad   T = (\size^\tab_1;\winner_1^\tab,\loser_1^\tab), \ldots, (\size^\tab_k;
            \winner_k^\tab, \loser_k^\tab))\,
                  \end{align*}
    \item For $\batchnum=1$ to $k$: {} 
      \begin{minipage}[t]{5cm} \label{step:correct sizes}
        \begin{enumerate}[leftmargin=4mm,nosep]
        \item $\size_\batchnum^\tab := \size_\batchnum^\act$; 
        \item $\winner_\batchnum^\tab := \min(\winner_\batchnum^\tab,\size_\batchnum^\act)$;\label{step:correct winner}
        \item $\loser_\batchnum^\tab := \min(\loser_\batchnum^\tab,\size_\batchnum^\act)$.\label{step:correct loser}
        \end{enumerate}
        \end{minipage}
    \item \label{step:compute margin} Let $S^\act, S^\tab  := \sum_{\batchnum=1}^k \size^\tab_\batchnum = \sum_{\batchnum=1}^k \size^\act_\batchnum$ and \[ \mu := \frac{\sum_{\beta=1}^k(\winner^\tab_\beta - \loser^\tab_\beta)}{\size^\act}.\]
      \label{step:correct tabulation}
    \item If $\mu \leq 0$ return $\mathtt{Inconclusive}$.
    \item Initialize $\iter=0$.  
    \item Repeat
      \begin{enumerate}[leftmargin=1cm,nosep]
      \item Increment $\iter := \iter + 1$.
      \item Perform $\disc_{\iter} := \mathtt{BasicExperiment}$
      \end{enumerate}
      until $\Stopmu(\disc_1,..., \disc_\iter)=1$
    \item If $\Criterionmu(\disc_1,..., \disc_\iter)=1$ return $\mathtt{Consistent}$; otherwise return $\mathtt{Inconclusive}$.
    \end{enumerate}
    
\underline{$\mathtt{BasicExperiment}$}:
    \begin{enumerate}[noitemsep]
    \item Select batch $\batchnum$ with probability $\size^\tab_\batchnum/ \size^\tab$.
    \item \label{step:start batch} Request CVR for batch $\batchnum$.  Denote the response
      $\cvr_\batchnum$.
    \item Apply $\cvrtransform$: $\cvr_\batchnum := \cvrtransform(\size_E^\act, T, \cvr_{\batchnum})$.
    \item $\mathtt{RowSelect}$: Select a row $r \in [\size^\tab_\batchnum]$ uniformly.
    \item If $\mathtt{CheckConsistent}(\size_E^\act, T, \cvr_\batchnum) = \mathtt{Error}$, return $2$.
    \item       Let $\iota$ be the ballot identifier in row $r$;\label{step:start ballot}
  request delivery of ballot $\iota$ from batch $\batchnum$.  
\item If a ballot $\mathbf{b}$ is delivered from batch $\beta$ with
  identifier $\iota$, let $\winner^\act, \loser^\act$ denote the
  $\{0,1\}$ values on $\mathbf{b}$ for the declared winner and loser
  respectively. Otherwise, set $\winner^\act := 0, \loser^\act := 1$. \label{step:ballot id check}
    \item Return $(\winner_r^\cvr -
    \loser^\cvr_r) - (\winner^\act - \loser^\act)$. \label{step:check consistent}
    \label{step:stop ballot}
    \end{enumerate}

\underline{$\mathtt{CheckConsistent}(\size_E^\act, T, \cvr_\batchnum)$}:
\begin{enumerate}[noitemsep]
\item If $\cvr_\batchnum$ is not uniquely-labeled 
  (Def.~\ref{def:cvr format}) return $\mathtt{Error}$.
\item If $\size^\cvr_\batchnum \neq \size^\act_\batchnum$ or $\size^\act_\batchnum \neq \size^\tab_\batchnum$, return $\mathtt{Error}$.
\item If $\winner^\cvr\neq \winner^\tab$ or $\loser^\cvr \neq \loser^\tab$, return $\mathtt{Error}$.
\item Return $\mathtt{OK}$.
\end{enumerate}
\vspace{-.1in}
\end{framed}
\vspace{-.1in}
\caption{The auditor $\Auditor[\cvrtransform;(\Stop, \Criterion)]$. Here $\cvrtransform$ is a CVR transform and $(\Stop, \Criterion)$ is an adaptive audit test.}
\label{fig:adaptive-auditor}
\end{figure}

\begin{figure}[t]
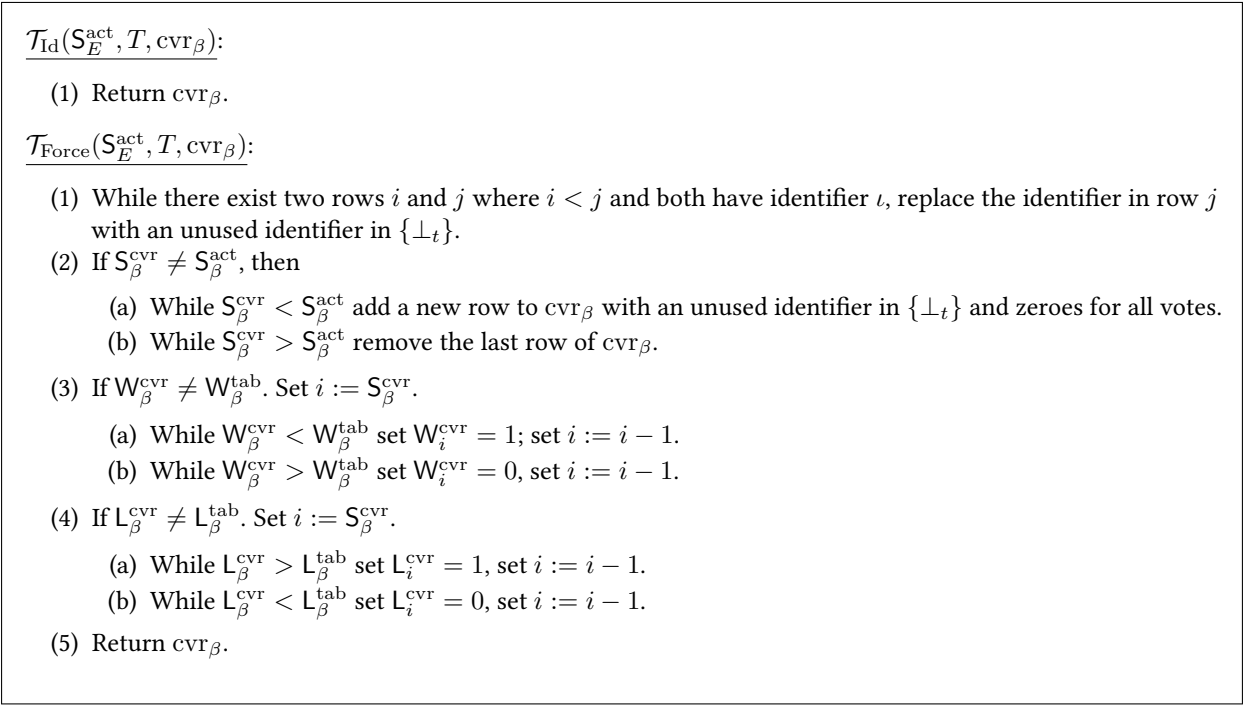

  \begin{framed}
\underline{$\cvrtransform_{\mathrm{Id}}(\size_E^\act, T, \cvr_\batchnum)$}:
\begin{enumerate}
\item Return $\cvr_\batchnum$.
\end{enumerate}

\underline{$\cvrtransform_{\mathrm{Force}}(\size_E^\act, T, \cvr_\batchnum)$}:
\begin{enumerate}[noitemsep]
\item \label{step:unique id} While there exist two rows $i$ and $j$
  where $i<j$ and both have identifier $\iota$, replace the identifier
  in row $j$ with an unused identifier in $\{\badrow_t\}$.
\item \label{step:sizes consistent} If $\size_\batchnum^\cvr \neq \size_\batchnum^\act$, then 
\begin{enumerate}[noitemsep]
\item While $\size_\batchnum^\cvr< \size_\batchnum^\act$ add a new row
  to $\cvr_\batchnum$ with an unused identifier in $\{\badrow_t\}$ and
  zeroes for all votes.
\item While $\size_\batchnum^\cvr> \size_\batchnum^\act$ remove the last row of $\cvr_\batchnum$.  
\end{enumerate}
\item \label{step:winner consistent} If $\winner_\batchnum^\cvr \neq \winner_\batchnum^\tab$. Set $i := \size_\batchnum^\cvr$. 
\begin{enumerate}[noitemsep]
\item While $\winner_\batchnum^\cvr< \winner_\batchnum^\tab$ set $\winner_i^\cvr = 1$; set $i := i-1$.
\item While $\winner_\batchnum^\cvr> \winner_\batchnum^\tab$ set $\winner_i^\cvr=0$, set $i := i-1$.
\end{enumerate}
\item \label{step:loser consistent} If $\loser_\batchnum^\cvr\neq  \loser_\batchnum^\tab$. Set $i := \size_\batchnum^\cvr$.
\begin{enumerate}[noitemsep]
\item While $\loser_\batchnum^\cvr> \loser_\batchnum^\tab$ set $\loser_i^\cvr=1$, set $i := i-1$.
\item While $\loser_\batchnum^\cvr< \loser_\batchnum^\tab$ set $\loser_i^\cvr=0$, set $i := i-1$.
\end{enumerate}
\item Return $\cvr_\batchnum$.
\end{enumerate}
\end{framed}
\caption{CVR transform functions.}
\label{fig:transforms}
\end{figure}

Figure~\ref{fig:adaptive-auditor} distinguishes two important
algorithmic elements of the auditor by giving them separate
``modular'' treatment: the statistical test and the CVR transform.
\begin{enumerate}
\item The statistical test. The auditor requires an adaptive audit test $(\Stop, \Criterion)$ as defined
  in Definition~\ref{def:audit-test}.
\item The CVR transform. The auditor requires a CVR transform $\cvrtransform$, which is a rule for rewriting a CVR before comparison.
\end{enumerate}
Thus a full description of the auditor is written $\Auditor[\cvrtransform;(\Stop, \Criterion)]$. In situations where the transform or the test are not directly relevant or can be inferred from context, we simply write $\Auditor$.

\paragraph{Remarks on the auditor's handling of the CVR.} As a
convenience, our treatment permits the Auditor to carry out
bookkeeping using the CVR, such as adding new rows or relabeling
certain rows 
with new identifiers that are known not to
match a physical ballot. For this purpose, we treat
$\badrow_1, \badrow_2, \ldots$ as a sequence of special purpose
identifiers known not to match any ballot. These modifications are for
internal bookkeeping of the auditor only; the original CVR is still
considered an immutable artifact of the audit.

The CVR separately records, for a given row $r$, whether it is
associated with a vote for $\winner$ or a vote for $\loser$; this
convention permits, in principle, rows of the CVR to contain votes for
\emph{both} candidates, known as an overvote, (a row with $1\, \, 1$ in the CVR table). This does not
interfere with the risk limit of the auditor (even when used for an
election that forbids overvotes) and is convenient for the
Force transform.
We point out in Appendix~\ref{app:exclusive} that this is unnecessary,
presenting a more complicated auditor that does not allow overvotes and a more complicated CVR
transform function that never creates overvotes.


\subsubsection{The CVR transform}
The auditor also takes as input a CVR rewriting procedure, denoted
$\cvrtransform$, that will be used to ``correct'' the CVR before
deciding if it is consistent with the tabulation.  Our proof that the auditor is
risk-limiting adopts the ``identity'' $\cvrtransform$ that does not
rewrite the CVR.  In Lemma~\ref{lem:permissive okay}, we then show
that if $\Auditor[\cvrtransform_{\mathrm{Id}}; (\Stop, \Criterion)]$ is
risk-limiting for the identity transform then
it is risk-limiting for \emph{any} procedure $\cvrtransform'$.
The goal of $\cvrtransform_{\mathrm{Force}}$ is to make the CVR consistent with tabulation with minimal edits. We use $\cvrtransform_{\mathrm{Force}}$ in all of our
completeness analyses.

\subsection{Discussion; an intuitive survey of the adaptive auditor}
\label{sec:checks needed}
We prove the soundness of the auditor in Section~\ref{sec:soundness};
this informal discussion is for the sake of intuition.

The $\mathtt{CheckConsistent}$ procedure returns an error (resulting
in a discrepancy of $2$) in many settings that could occur naturally
in practice, such as a mismatch between the number of ballots counted
on the tabulator and the number of ballots on the CVR. 
Here we discuss the role played by the various properties checked by
$\mathtt{CheckConsistent}$. We remark again that a much more
permissive auditor is obtained by the Force transform, discussed
later.

\paragraph{Uniquely labeled CVRs.} In our model and in many practical settings the auditor cannot ensure that ballots are uniquely
labeled. This explains the convention that defines
discrepancy for a row $r$ as the minimum discrepancy across all
ballots with the row identifier $\iota_r$. The auditor does, however,
ensure the uniqueness of identifiers appearing in the CVR. A concrete attack
exists in the absence of this check.  One simply labels all ballots with the same identifier
and crafts a CVR to be consistent with tabulation.  Then when a ballot is requested
one simply returns a ballot with the votes listed in the CVR row.  This attack
succeeds as long as all vote patterns exist on at least one ballot. This is why a crucial
step in $\cvrtransform_{\mathrm{Force}}$ in Figure~\ref{fig:transforms} is to rewrite duplicate identifiers on a CVR. 

\paragraph{Treatment of missing ballots.} Missing ballots are
treated as though cast for the loser. If not, the adversary can always
choose to not return those ballots that show votes for the loser,
effectively reducing the observed discrepancy. This treatment is similar to the ``phantoms to zombies'' approach~\cite{banuelos2012limiting}.

\paragraph{Enforcing equality of batch sizes.} 
The size
checks ensure that the $\mathtt{RowSelect}$ operation selects both a uniform row in the CVR and (for an honest adversary) a uniform ballot in the batch.

\paragraph{Enforcing equality of CVR and tabulation subtotals.} As
discussed below, the tabulation effectively determines a lower bound
on total discrepancy for the batch regardless of adversarial choice of
the CVR.  Without the check $\winner^\cvr = \winner^\tab$ and
$\loser^\cvr = \loser^\tab$, the CVR could always be consistent with
the ballots without actually auditing the tabulation.

\section{An Adversarial Auditing model}
\label{sec:adversary model}
As discussed in Section~\ref{ssec:statistical tests}, the conventional
formal approach to RLAs adopts the language of Neyman--Pearson
statistical hypothesis testing.
%
%
This
picture emphasizes the role played by the culminating
statistical test. Our more complex setting---involving adaptive
selection of CVRs that may depend on the entire history of the
audit---motivates us to extend the formal treatment of the audit to
the entire procedure.
We adopt the ``security game'' framework from the theory of
cryptography, which has the expressive power to reflect such
interactions between parties. The cryptographic model has the
advantage that it explicitly identifies an \emph{adversary}, a party
that is charged with frustrating or subverting the audit, and
precisely defines which aspects of the audit are under adversarial
control.  

In our framework, the adversary is responsible for producing CVRs and
providing ballots to the auditor when requested; ballot labels are
also effectively under adversarial control, as the final conclusions
are guaranteed for all such labelings.
The resulting game is a ``physical cryptography game'' along
the lines of Fisch, Freund, and Naor~\cite{fisch2014physical}.  In
general, our definition gives the adversary control over parts of the
process whenever possible. This explicitly identifies what aspects of
the procedure must be honestly conducted for the statistical
guarantees to hold.  Finally, we remark that we adopt the classical
nomenclature of ``soundness'' and ``completeness'' for cryptographic
games that act as the analogues of Type I and Type II errors.

\paragraph{The Auditor--Adversary Game.}
The \emph{Auditor--Adversary} game is played by two parties, the
\emph{Auditor} denoted by $\Auditor$ and the \emph{Adversary} denoted
by $\Adversary$. The game is played in the context of an
election (Definition~\ref{def:election}) and involves the exchange of
both physical objects (ballots) and information (CVRs).
 Recall that we use boldface to refer to physical objects which may be exchanged between the formal parties in the game.


\label{ssec:formal game} Figure~\ref{fig:auditing game} describes in
detail the adaptive RLA game between the auditor and adversary. Before
discussing the desired risk and completeness properties, we discuss
our ballot identification convention.

\begin{figure}[t]
  \begin{framed}
  Auditor ($\Auditor$)--Adversary ($\Adversary$) game for 
    election $E = (\mathbf{B}, T)$
    \begin{enumerate}[noitemsep]
  \item \textbf{Setup}.
    \begin{enumerate}[noitemsep]
    \item\textbf{Ballot and tabulation delivery (to $\Adversary$).} The
      physical ballots $\mathbf{B}$ and the tabulation $T$ are given to the
      adversary $\Adversary$.
    \item\textbf{Ballot manifest and tabulation delivery
        (to $\Auditor$).}  The ballot manifest
      $\size_E = (\size_1^\act, \ldots, \size_k^\act)$ and the tabulation $T$ are
      given to the auditor $\Auditor$.
    \end{enumerate}
  \item \textbf{Audit}. $\Auditor$ repeatedly makes one of the
    following two requests of $\Adversary$, or chooses to conclude the
    audit:
    \begin{itemize}[noitemsep]
    \item \textbf{A CVR request}. $\Auditor$ requests a CVR for
      batch $\batchnum$.
      $\Adversary$
        responds with a CVR denoted $\CVR_\batchnum$.
    \item \textbf{A ballot request}. $\Auditor$ requests a ballot
      from the adversary with a specific identifier
      $\iota \in \{0,1\}^*$ from some batch $\batchnum$.
      \begin{enumerate}[noitemsep]
      \item $\Adversary$ either
      sends a physical ballot $\mathbf{b}$ in batch  $\batchnum$, i.e. $\mathbf{b} \in \mathbf{B}_\batchnum$, to $\Auditor$ or responds with \textbf{No
      ballot.}
    \end{enumerate}
    \end{itemize}
  \item \textbf{Conclusion}. $\Auditor$ returns one of the two values:
    \begin{description}[noitemsep]
      \item[$\consistent$] meaning ``Audit consistent with tabulation,'' or
      \item[$\inconclusive$] meaning ``Audit inconclusive.''
        \end{description}
\end{enumerate}
\end{framed}
\vspace{-.1in}
\caption{The $\RLA_{\Auditor,\Adversary}(E)$ auditing game.}
\vspace{-.1in}
\label{fig:auditing game}
\end{figure}

\paragraph{Ballot identification.} 
\label{step:ballot id}
Our definition of a ballot family (Definition~\ref{def:ballot-family})
includes identifiers on ballots. Recall that ballot identifiers are
not assumed to be unique, which reflects an important feature of
practical RLAs: in general, it's not possible for auditors to
efficiently check physical identifiers to ensure that there are no
collisions.

Our results work perfectly well if the adversary is permitted to
(re-)assign identifiers to a batch each time they are asked to
generate a CVR for that batch (this may be the case if a tabulator imprints during the audit). There are two crucial assumptions
required for security in this setting: \begin{enumerate*} \item the adversary cannot
change ballot identifiers unless another CVR is requested for the
batch, and \item the auditor---if ever given the chance to observe the
ballot---can reliably determine $\id_{\mathbf{b}}$. \end{enumerate*}

An adversary can effectively ``destroy'' a ballot by choosing not
to reveal it when requested.

\begin{definition}[Risk; soundness]
  Let $\Auditor$ be an Auditor.  For election $E$ and adversary 
  $\Adversary$ let
  $\RLA_{\Auditor,\Adversary}(E)$ denote the random variable
  equal to the conclusion of the audit as described in Figure~\ref{fig:auditing game}.
An auditor $\Auditor$ has \emph{$\risk$-risk} (or
    \emph{$\risk$-soundness}) if, for all invalid elections $E$ and all
    adversaries $\Adversary$,    \label{def:risk}
\[
      \Pr[\RLA_{\Auditor,\Adversary}(E) = \consistent] \leq \risk\,.
    \]
    (The probability here is taken over random choices of the auditor and the adversary.)
\end{definition}

\noindent
Definition~\ref{def:risk} is a property of a $\Auditor$ (the
auditor) only.  That is, it holds for all invalid elections and behaviors of the adversary. As we discuss in Section~\ref{sec:correctness}
completeness or Type-II errors will only be guaranteed for certain
$\Adversary$.



\subsection{Modeling conventional RLAs} \label{ssec:conventional rlas}This modeling can apply directly to conventional ballot-comparison audits. In particular, by restricting the class of adversaries to those that draw all batch CVRs from a fixed global CVR, one obtains a model that corresponds to a conventional comparison audit. In particular, as this is a smaller class of adversaries, all of the conclusions of the paper apply to this setting (including the conclusions for the specific auditor we consider). This auditor can provide privacy improvements over traditional auditors, as it only needs to release portions of the global CVR table. As an alternate modeling approach, one can formulate an auditor that initially requests the entire CVR; with this convention, one can return to universally quantifying over all adversaries. The risk limits for this auditor follow directly from our proofs. Finally, we mention that these techniques demonstrate that traditional RLAs do not require the uniqueness of physical ballot identifiers.  

The model can also be adapted to reason about polling audits, where
auditors never issue CVR requests and tacitly assume a ``position based'' labeling. For simplicity, this variant calls for the adversary to label all ballots at the outset. These labels are never communicated to the auditor, who simply assumes that ballots are given labels of the form $(b,n)$, where $b$ is a batch number and $n$ is a ``sequence number'' between $1$ and the size of the batch. (Note that the auditor can deduce this label set from the ballot manifest.) Intuitively, this corresponds to the natural setting where ballots in each batch are placed in order and selection is determined by identifying a particular index in a particular batch.
We remark that there are ballot polling techniques that are not
directly reflected by this modeling: for example, techniques that
treat ``asking for a random ballot'' as an atomic operation. (For example \emph{$k$-cut} which cuts a stack of
ballots an appropriate number of times~\cite{sridhar2020k}.) Of course, with further alterations to the model, this could also be treated as a (necessarily) trusted operation.

\section{$\Auditor[\cvrtransform_{\mathrm{Id}}; (\Stop, \Criterion)]$ is Risk-Limiting} 
\label{sec:soundness}
 The key for establishing that $\Auditor$ is risk-limiting is to demand that the generated CVR is nearly
  consistent with the previously generated tabulation. We observe that
  with this assurance, the tabulated results effectively generate a
  forcing ``commitment'' on the discrepancy
  of any CVR that the adversary may
  generate. Batch tabulations now play an essential role in the
  analysis by enforcing this commitment. In a conventional ballot comparison audit, the details of
  the tabulation itself can be ignored so long as the tabulation and
  CVR declare the same winner: The operational details
  of the audit are determined entirely by the CVR.

\begin{theorem}
\label{thm:main}
Let $(\Stop, \Criterion)$ be an adaptive audit test with risk
$\alpha$; let $\cvrtransform$ be an arbitrary procedure that
transforms CVRs to CVRs. Let $\Auditor$ the auditor in
Figure~\ref{fig:adaptive-auditor}. Then
$\Auditor[\cvrtransform;(\Stop, \Criterion)]$ has risk $\alpha$.
\end{theorem}

\begin{proof}
  We begin with the next Lemma, showing that a $\cvrtransform$ does
  not affect whether an auditor is risk-limiting.
\label{ssec:cvr transform}

\begin{lemma}
\label{lem:permissive okay}
Let $\cvrtransform$ be a (possibly randomized) procedure that takes as
input $(\size_E^\act, T, \cvr_\batchnum)$ and rewrites
$\cvr_\batchnum$.  Let $(\Stop, \Criterion)$ be a statistical test and
let $\Auditor$ be an auditor as in Figure~\ref{fig:adaptive-auditor}.

If $\Auditor[\cvrtransform_{\mathrm{Id}}; (\Stop, \Criterion)]$ satisfies Definition~\ref{def:risk} with $\risk$-risk then $\Auditor[\cvrtransform; (\Stop, \Criterion)]$ satisfies Definition~\ref{def:risk} with $\risk$-risk.
\end{lemma}

\noindent
The proof of Lemma~\ref{lem:permissive okay} has a simple core: For every adversary, $\Adversary$ that succeeds in the presence of $\cvrtransform$ one can define another adversary $\Adversary'$ that applies $\cvrtransform$ before returning the CVR to the auditor. 

\begin{proof}
We show the result by the contrapositive.  Fix some statistical test $(\Stop, \Criterion)$. Suppose that for some election $E$ there exists an adversary $\mathcal{A}$ such that
    \[
      \Pr_{\mathcal{C}_{\cvrtransform, (\Stop, \Criterion)}}[\RLA_{\mathcal{C}_{\cvrtransform, (\Stop, \Criterion)},\mathcal{A}}(E) = \consistent] >\risk\,.
    \]

Consider $\Auditor[\cvrtransform_{\mathrm{Id}};(\Stop, \Criterion)]$.  
Assume for a moment that the test $\Stop$ always outputs $0$. (This is just to define a sequence of length $\ell$, noting that the selection of batches/ballots is independent in each iteration though the resulting discrepancies need not be independent).

Fix some positive number $\ell$ and consider a sequence of selected batches $\batchnum_1,..., \batchnum_\ell$ and selected locations within a batch $\iota_1,..., \iota_\batchnum$ with $\iota_\batchnum = \perp$ as a special value indicating that no ballot is selected. Here we that note both of these sequences of random variables are independent of an adversary and only depend on the election $E$.  Furthermore, note that these sequences are identically distributed in $\Auditor[\cvrtransform_{\mathrm{Id}};(\Stop, \Criterion)]$  and $\Auditor[\cvrtransform;(\Stop, \Criterion)]$  except that some locations may be $\perp$ in either sequence but not in the other.
Consider the following adversary $\mathcal{A}'$ for the auditing experiment with $\Auditor[\cvrtransform_{\mathrm{Id}};(\Stop, \Criterion)]$. 
\begin{itemize} 
\item $\mathcal{A}'$ initializes $\mathcal{A}$ with $E$.  
\item $\mathcal{A}'$ runs $\mathcal{A}$ and forwards all audit requests  to $\mathcal{A}$.  Upon receiving a response $\cvr_{\batchnum}$ from $\mathcal{A}$, compute
$
    \cvr_\batchnum' = \cvrtransform(\size_E^\act, T, \cvr_\batchnum)
  $
  and return $\cvr'_\batchnum$ to $\Auditor[\cvrtransform_{\mathrm{Id}};(\Stop, \Criterion)]$.  
\item Upon receiving request for ballot $\iota$, forward request to $\mathcal{A}$ and return ballot returned by $\mathcal{A}$.
\end{itemize}

\noindent
$\mathcal{A}'$ exactly replicates the view that $\mathcal{A}$ would experience interacting with $\Auditor[\cvrtransform;(\Stop, \Criterion)]$. The sequence of batches and locations selected in $\Auditor[\cvrtransform_{\mathrm{Id}};(\Stop, \Criterion)]$ when interacting with $\mathcal{A}'$ is identically distributed to $\Auditor[\cvrtransform;(\Stop, \Criterion)]$ when interacting with $\mathcal{A}$. 

We define
$\vec{\disc}_{\Auditor[\cvrtransform;(\Stop, \Criterion)],
  \mathcal{A}}$ as the sequence of discrepancies produced by
$\mathcal{A}$ when interacting with $\Auditor[\cvrtransform;(\Stop, \Criterion)]$.  Similarly, define
$\vec{\disc}_{\Auditor[\cvrtransform_{\mathrm{Id}};(\Stop, \Criterion)], \mathcal{A}'}$ as the sequence of discrepancies
produced by $\mathcal{A}'$ interacting $\Auditor[\cvrtransform_{\mathrm{Id}};(\Stop, \Criterion)]$.  We now remove the assumption that $\Stop$ always outputs $0$. Then, the
two sequences
$\vec{\disc}_{\Auditor[\cvrtransform;(\Stop, \Criterion)],
  \mathcal{A}}$ and
$\vec{\disc}_{\Auditor[\cvrtransform_{\mathrm{Id}};(\Stop, \Criterion)], \mathcal{A}'}$ are identically distributed.  Thus,
it must be the case that
    \[
      \Pr_{\mathcal{C}}[\RLA_{\vec{\disc}_{\Auditor[\cvrtransform_{\mathrm{Id}};(\Stop, \Criterion)], \mathcal{A}'}}(E) = \consistent] >\risk\,.
    \]
This is a contradiction and proves Lemma~\ref{lem:permissive okay}.
\end{proof}

We then analyze $\mathtt{BasicExperiment}$ defined in
Figure~\ref{fig:adaptive-auditor} where a batch is selected with
probability proportional to its actual size and a uniform row is
selected from the generated CVR. Before analyzing a single iteration of $\mathtt{BasicExperiment}$,
we consider the result of Steps~\ref{step:start batch} to
\ref{step:stop ballot} in $\mathtt{BasicExperiment}$ for some fixed
$\batchnum$ and adversary $\Adversary$ (and the identity CVR
transform). That is, we focus on the random variables $r$ and
$\obsdisc_{\batchnum}$ defined by the following procedure and denoted as $\mathtt{BasicExperiment}_\batchnum$.

\smallskip
\noindent \emph{Definition of the random variables $r$ and
  $\obsdiscbatch$}:
\begin{enumerate}
\itemsep0em
\item $\Adversary$ generates a CVR for $\batchnum$, denoted $\cvr$.
\item A row $r \in [\size^\man_b]$ is drawn independently and
  uniformly at random.
\item $\obsdiscbatch$ is defined to be $2$ if $\mathtt{CheckConsistent}$ outputs $\mathtt{Error}$.
\item If $\obsdiscbatch$ has not already been set to $2$ in the
  step above, let $\iota$ be the identifier appearing in row $r$. The
  adversary is asked to return a ballot from batch $\batchnum$ with
  identifier $\iota$. If the adversary responds with such a ballot
  $\mathbf{b}$,
  $\obsdiscbatch = (\winner_r - \loser_r) - (\winner_{\mathbf{b}}
  - \loser_{\mathbf{b}})$; otherwise
  $\obsdiscbatch = (\winner_r - \loser_r) + 1$.
\end{enumerate}

\begin{claim}
\label{clm:local}
Consider $\mathtt{BasicExperiment}_\batchnum$ for an adversary $\mathcal{A}$ and a batch $\batchnum$. Then
  $
    \Exp[\obsdiscbatch] \geq \disc_\batchnum/\size_\batchnum\,.
  $
\end{claim}

\begin{proof}
  The random variable $\obsdisc_\batchnum$ is determined by selection of
  $\cvr$ by $\Adversary$, (independent) uniform selection of $r$ by
  $\Auditor$, and final selection by $\Adversary$ of a ballot to
  return.
  The proof only requires that $\cvr$ and $r$ are independent; in
  particular, $\cvr$ may be chosen with arbitrary dependence on the
  history of the audit.
  We remark that the same guarantee holds if multiple instances of
  $\mathtt{BasicExperiment}_\batchnum$ occur in parallel, as the independence
  assumption is guaranteed by $\Auditor$.

  We will show that the inequality holds conditioned on any fixed CVR
  $\cvr$ produced by the adversary in the first step; hence
  it holds for any distribution of CVRs. Note that if
  $\mathtt{CheckConsistent} = \mathtt{Error}$ for this CVR then
  $\obsdisc_\batchnum = 2$ and the claim is clearly true. Otherwise,
  $\mathtt{CheckConsistent} = \mathtt{OK}$, the CVR
  $\cvr =
  ((\iota_1,\winner_1,\loser_1),\ldots,(\iota_s,\winner_s,\loser_s))$
  is uniquely-labeled, $s = \size^\cvr_\batchnum = \size^\man_\batchnum$,
  $\winner^\cvr_\batchnum = \winner^\tab_\batchnum$, and
  $\loser^\cvr_\batchnum = \loser^\tab_\batchnum$.

  For any particular row $r$ of the $\cvr$, let
$
    \mathbf{B}(r) = \{ \mathbf{b} \in \mathbf{B}_\batchnum \mid
    \iota_\mathbf{b} = \iota_{r}\}
  $
  denote the set of ballots with identifier that matches $\iota_{r}$.
  Consider the following function of ballots in batch $\batchnum$,
  denoted
  $\match:[\size_\batchnum] \rightarrow
  \mathbf{B}_\batchnum$
  to rows in the CVR:
  \begin{enumerate}
    \itemsep0em
  \item For a row $r$ for which $|\mathbf{B}(r)| \geq 1$ associate any
    ballot $\mathbf{b} \in \mathbf{B}(r)$ with $r$ that minimizes the
    resulting discrepancy (and hence achieves $\disc_r$ from Definition~\ref{def:discrepancy}).
  \item Of the remaining, yet unassociated, ballots, assign them
    arbitrarily, but in a one-to-one fashion, to the rows of the CVR
    which have ballot identifiers that do not match a physical ballot.
  \end{enumerate}
  As the CVR is uniquely-labeled there is no contention for the ballots assigned by the first
  rule.  That is, $\match$ is a one-to-one function between rows and
  physical ballots.  Furthermore, since
  $\size^\act_\batchnum = \size^\cvr_{\batchnum}$ the function
  $\match$ is also onto; thus $\match$ is bijective.

  For this fixed $\batchnum$ and fixed $\cvr$ provided by
  $\Adversary$, let $\corrdisc$ denote the random variable (determined
  by the random variable $r$) given by the discrepancy between the
  votes appearing in row $r$ and $\match(r)$.  That is,
  \[\corrdisc= (\winner_r - \loser_r) - (\winner_\mathbf{b} -
    \loser_\mathbf{b}).\] We then note that, conditioned on observing
  a fixed $\cvr$,
  \[
    \corrdisc {\leq}^{(1)}\; \disc_r^{\cvr} {\leq}^{(2)}\; \obsdiscbatch
  \]
  with certainty over the uniform choice of $r$.
  
  The inequality ${\leq}^{(1)}$ follows immediately from the
  definition of $\disc_r^{\cvr}$: to see this, observe that if
  $\mathbf{B}(r) \geq 1$ then there is a matching ballot and
  $\corrdisc = \disc_r^{\cvr}$ as they are both determined by
  minimum discrepancy obtained over all matching ballots; if, on the other
  hand, there is no matching ballot then the inequality follows
  because
  $(\winner_r - \loser_r) - (\winner_\ballot - \loser_\ballot) \leq
  (\winner_r - \loser_r) + 1$ for any ballot $\mathbf{b}$. 
  
  As for the second inequality ${\leq}^{(2)}$, note that if
  the adversary returns a ballot that matches the identifier for row
  $r$, $\disc_r^{\cvr} \leq \obsdiscbatch$ as above, since
  $\disc_r$ is defined to be the minimum value over all matching
  ballots. If the adversary does not return a matching ballot then
  $\disc_r \leq (\winner_r - \loser_r) + 1 = \obsdisc_\batchnum$,
  as desired.

  We conclude that
  \begin{equation}\label{eq:conditional}
    \Exp\left[\obsdisc_\batchnum\right] = \sum_{\cvr} \Pr[\text{$\Adversary$ generates $\cvr$}] \Exp[\obsdisc_\batchnum\mid \cvr]\geq \sum_{\cvr} \Pr[\text{$\Adversary$ generates $\cvr$}] \Exp_r\left[\corrdisc\right]\,.
  \end{equation}
  For a fixed $\cvr$, we may expand $\Exp\left[\corrdisc\right]$ as the sum
  \begin{equation}\label{eq:corrdisc}
\frac{1}{\size_\batchnum} \sum_{r = 1}^{\size_\batchnum}
                       \left( (\winner^\cvr_r - \loser^\cvr_r) - (\winner_{\match(r)}^\act - \loser_{\match(r)}^\act)\right)\,.
                     \end{equation}
 As $\match$ is bijective, every ballot appears exactly once in this sum, so we can rewrite the quantity in \eqref{eq:corrdisc}
                     \[
                       \frac{1}{\size_\batchnum} \left(\sum_R (\winner^\cvr_R - \loser^\cvr_R) - \sum_{\ballot \in \mathbf{B}_\batchnum}(\winner_\ballot^\act - \loser_\ballot^\act)\right)= \frac{\disc_\batchnum}{\size_\batchnum}\,.
                     \]
                     Returning to~\eqref{eq:conditional}, we have
  \begin{equation}
    \begin{split}
      \Exp\left[\obsdisc_\batchnum\right] &\geq \sum_{\cvr} \Pr[\text{$\Adversary$ generates $\cvr$}] \Exp\left[\corrdisc\right]\\
      &= \sum_{\cvr} \Pr[\text{$\Adversary$ generates $\cvr$}] \frac{\disc_\beta}{\size_\beta}\\
      &= \frac{\disc_\beta}{\size_\beta} \sum_{\cvr} \Pr[\text{$\Adversary$ generates $\cvr$}] =\frac{\disc_\beta}{\size_\beta}\,,
    \end{split}\nonumber
  \end{equation}
  which completes the proof of Claim~\ref{clm:local}.
\end{proof}

We now turn to analyzing a single iteration of
$\mathtt{BasicExperiment}$. We define the result of
this experiment to be a random variable $\obsdisc$,
defined by the following procedure:
\begin{enumerate}
\item Select a batch $\batchnum$ with probability $\size^\man_\batchnum/\size^\man$.
\item Carry out the local experiment with batch $\batchnum$. 
\end{enumerate}

\begin{claim} \label{clm:basic} For any adversary $\Adversary$, the
  expectation of $\obsdisc$ over a single iteration satisfies
  \begin{align*}
    \Exp[\obsdisc] &= \sum_\batchnum \left(\frac{\size^\man_{\batchnum}}{\size^\man} \cdot
                     \Exp[\obsdiscbatch] \right) 
                     \geq \sum_\batchnum \left(\frac{\size^\man_\batchnum}{\size^\man} \cdot
                     \frac{\disc_\batchnum}{\size_\batchnum^\man}\right) = \frac{\disc}{\size}\,.
  \end{align*}
  \label{clm:one iter}
\end{claim}

\noindent
Theorem~\ref{thm:main}  follows from Claim~\ref{clm:one iter} by noting that for any invalid election the input $\obsdisc$ to $(\Stop, \Criterion)$ is a $\disc/\size \ge \mu^\tab$
dominated random variable and by application of Lemma~\ref{lem:permissive okay}.
\end{proof}

\subsection{Concrete statistical tests}
\label{ssec:concrete tests}

We recall the Kaplan-Markov test.

\begin{definition}[Kaplan-Markov \cite{stark2009efficient,stark2009auditing,stark2010super,Stark:Conservative}]
  Let $\alpha\in [0,1]$, $\gamma>1$,
  $\ell_{\min}, \ell_{\max} \in \mathbb{Z}^+$.  Define the value
  \[ \Riskdelta^{(\gamma)}(\disc_1,..., \disc_\ell ) =
    \prod_{\iter=1}^\ell
    \left(\frac{1-\frac{\delta}{2\gamma}}{1-\frac{\disc_\iter}{2\gamma}}\right)\,.
\]
The $(\alpha, \gamma, \ell_{\min}, \ell_{\max})$-\emph{Kaplan-Markov} audit statistical test is $(\Stop, \Criterion)$ where 
$\Stopdelta(\disc_1,..., \disc_\ell) =1$ if and only if
\[
  \ell\ge \ell_{\max} \vee \left(\Riskdelta^{(\gamma)}(\disc_1,..., \disc_\ell, ) \le \alpha \wedge \ell\ge \ell_{\min}\right)\quad\text{and}
  \quad \Criteriondelta(\disc_1,..., \disc_\ell, ; \gamma)  = \left(\Riskdelta^{(\gamma)}(\disc_1,..., \disc_\ell, ; \gamma) \le \alpha\right).
  \]
\label{def:km}
\end{definition}

\noindent
\textbf{Note:} One can define the test without $\ell_{\min}$ or $\ell_{\max}$.  The parameter $\ell_{\max}$ is usually set to some small fraction of the overall number of ballots where hand counting becomes more efficient.  The parameter $\ell_{\min}$ is usually set so that some number of sampled ballots can display $1$-vote overstatements while meeting the risk limit.  For a $\lambda* \delta$ fraction of $1$-vote overstatements to be acceptable
\[
  \ell_{\min} = -\log \alpha/\left(\delta\left(\frac{1}{2\gamma}+\lambda\log\left(1-\frac{1}{2\gamma}\right)\right)\right)
\]
suffices~\cite{stark2009efficient}. 

\begin{claim}\label{clm:km monotone}
  The Kaplan-Markov test is an adaptive audit test.
\end{claim}

\begin{proof}[Proof of Claim~\ref{clm:km monotone}]
  Consider a sequence of bounded, non-negative and i.i.d.\ real-valued
  random variables $X_1, \ldots$, each with mean $\delta$.  The
  Kaplan--Markov inequality asserts that
  \begin{equation}\label{eq:Kaplan-Markov}
    \Pr\left[\max_{t=0}^n \prod_{i = 1}^t (X_i/\delta) \geq 1/\alpha\right] \leq
    \alpha \qquad \text{for any $\alpha > 0$.}
  \end{equation}
  Critically, we observe that the Kaplan-Markov inequality applies to
  random variables under the weaker $\delta$-dominating
  condition. Specifically, assume that $X_1, X_2, \ldots$ are
  $\delta$-dominating (but not necessarily i.i.d.). Then the sequence of
  random variables $Z_1 = X_1/\delta, Z_2 = (X_1/\delta) (X_2/\delta), \ldots$
  form a nonnegative sub-martingale, which is to say that
  $\Exp[Z_t | Z_1, \ldots, Z_{t-1}] \geq Z_{t-1}$. According to the
  Doob (sub-)martingale inequality,
  $\Exp[\max_{i=1}^n Z_i] \leq \Exp[Z_n]$ and hence Markov's
  inequality can be applied to yield~\eqref{eq:Kaplan-Markov}, as
  desired. (See, e.g., \cite[\S14.6]{Williams:1991wk} for a detailed
  account of the Doob inequality). Finally, the Kaplan-Markov test for $\delta$-dominated random
  variables is obtained by applying~\eqref{eq:Kaplan-Markov} to the
  observed discrepancies under the transformation
  $D \mapsto 1 - D/(2 \gamma)$.
\end{proof}

Other classical tail bounds directly yield adaptive audit tests by
monotonicity or stochastic domination arguments. For example, the
Azuma-Hoeffding inequality applies to this situation as it applies
directly to submartingales. (See, e.g., \cite{Motwani:1995wb} for a
detailed account.) Inequalities that optimize one side of the tail
bound (e.g., the upper Chernoff bound) can be applied to this
situation via a stochastic dominance argument that exploits the fact
that the test criteria are monotone.


\section{Completeness}
\label{sec:correctness}

The second natural figure of merit for an audit is the probability
that it correctly concludes that a valid election is ``Consistent.''
Treating this issue is complicated by the fact that inconsistencies
between the CVR and the physical ballots are frequently observed
even during vigilant audits of valid elections. Thus, the underlying
statistical tests must be parameterized in order to tolerate a certain
frequency of errors. Ultimately, this leads to a trade-off between
risk, sample size, and the probability that a valid election will be
found inconclusive when the audit is subject to some presumed rate of
inconsistencies. This third quantity we call ``completeness''; this is
non-standard terminology motivated by directly analogous definitions
in cryptography.

The traditional analysis of completeness focuses on the number of
overstatements and understatements, either according to the actual
ballot population or observed empirically during the audit. The
relationship to sample size and risk then depends largely on the
details of the adopted statistical test (see \cite{nealcode,starkcode}
and Section~\ref{ssec:concrete tests}).
However, our setting introduces new types of inconsistencies that may
arise during an audit: in particular, mismatches between the
tabulation and CVR yield a new source of non-zero observed
discrepancy. 

To provide a comprehensive treatment, we augment the
traditional accounting of under- and overstatement errors with two
further classes of errors.  
\textbf{Ballot Additions} can result from
ballots that are scanned or tabulated more than once (which a
tabulator cannot detect without an identifier). \textbf{Ballot
  Deletions} can result from ballots that were cast but never scanned
or whose interpretations were not included in the reported results. We
remark such errors can also arise in traditional settings.  15\% of
audited precincts in Connecticut in the 2020 presidential election
reported a different ballot count from the
tabulation~\cite{russell2021statistical}.
To the best of our knowledge, this is the first formal detailed
analysis of the effect of additions and deletions.

\paragraph{Handling size, tally, and uniquely-labeled failures via the CVR transform mechanism.} Recall that the strict ``default'' auditor (that is, the procedure of Figure~\ref{fig:adaptive-auditor} using
$\cvrtransform_{\mathrm{Id}}$) rejects CVRs resulting from commonplace
errors. For example, if the CVR has one fewer row than the size of the
batch or if 
$\winner^\cvr=\winner^\tab+1$.
%
To eliminate such errors, $\cvrtransform_{\mathrm{Force}}$ forcibly
revises the CVR so as to declare sizes and vote totals consistent with
the manifest and tabulation. While this transformation corrects the
CVR in this sense, it may generate new overstatements or
understatements. The CVR transform paradigm provides a unified
way to treat such errors by converting them into understatement and
overstatement errors, which have a well understood effect on standard
statistical tests.

In light of the discussion above, this section provides precise
control on the effect of size mismatches, vote tally disagreements, or
duplicated identifiers on the resulting number of overstatements and
understatements.
%
%
With these equivalencies in hand, one can compute appropriate sample
sizes for different statistical tests by established
techniques~\cite{nealcode,starkcode}. As remarked above, this approach
can also be used to treat similar issues in traditional comparison
audits.
%
%

We separately present and analyze two different
settings.  The first setting considers a consistent CVR and
tabulation that disagree with the physical ballots. The second setting
considers an arbitrary tabulation in context of an inconsistent
CVR.
We compose these in Section~\ref{ssec:completeness discussion} to handle the  general case.

\begin{definition}[The canonical CVR]
\label{def:canonical CVR}
Let $\mathbf{B}$ be a uniquely labeled ballot
family.
%
A global CVR $\cvr^* = (\cvr^*_1, \ldots, \cvr^*_k)$ is
\emph{canonical} if it  correctly reflects the ballots. That is,
the ballots $\mathbf{B}_\batchnum$ can be placed in one-to-one
correspondence with the rows of $\cvr_\batchnum$ in such a way that
both the identifiers and votes match. 
For the ballot family $\mathbf{B}$, 
$\cvr^*_{\mathbf{B}}$ indicates a canonical
CVR.   
\end{definition}

\noindent
Observe that any canonical CVR is uniquely labeled.  The canonical CVR is only determined
up to a permutation of the rows. Despite this, we say ``the
canonical CVR'' of a ballot family.


\begin{definition}[The honest adversary]
  Let $E=(\mathbf{B},T)$ be an election with
  uniquely labeled ballots and let $\cvr$ be a uniquely labeled global CVR.  The
  \emph{honest adversary}
  $\mathcal{H}(\mathbf{B}, \cvr, T)$ is the
  adversary that
  responds to any CVR request with the appropriate $\cvr_i$ and
  responds to any request for an (existing) ballot identifier $\iota$
  with the matching ballot $\mathbf{b}$.  If no ballot exists matching
  the identifier, it returns \textbf{No ballot}.
  \label{def:canonical adversary}
\end{definition}

The honest adversary's behavior is only defined if all ballots
have unique identifiers and the $\cvr$ is uniquely labeled.



\begin{definition}[Pairwise CVR discrepancy] Let $\cvr_1, \cvr_2$ be
  two uniquely labeled CVRs (for the same batch of a ballot family).  For
  an identifier $\iota$ that appears in both CVRs, define
  \[ \disc(\cvr_1, \cvr_2, \iota) =(\winner^{\cvr_1}_{r_\iota} -
  \loser^{\cvr_1}_{r_\iota}) - (\winner^{\cvr_2}_{r_\iota} -
  \loser^{\cvr_2}_{r_\iota})\,.
  \]
\end{definition}
\begin{definition}[CVR distortion]
  Let $\mathbf{B}$ be a uniquely labeled ballot family and
  $\cvr = (\cvr_1, \ldots, \cvr_k)$ be a global CVR for $\mathbf{B}$.
  Let $(o_1, o_2, u_1, u_2, a, d)$ be natural numbers such that
  $\winner^{\cvr} , \loser^{\cvr}, \size^{\cvr}-\winner^{\cvr},
  \size^{\cvr}-\loser^{\cvr}$ are all at least $o_1+o_2+u_1+u_2+a+d$.
  Then $\tilde{\cvr}$ is a \emph{$(o_1, o_2, u_1, u_2, a, d)$-distortion} of
  $\cvr$ if $\tilde{\cvr} = \cvr$ with the following exceptions:
  \begin{description}
  \itemsep0em
  \item[Overstatements/Understatements.] There are 
  \begin{itemize}
  \item $o_1$ identifiers $\iota$ where $\disc(\tilde{\cvr}, \cvr, \iota)=1$,
  \item  $o_2$ identifiers $\iota$ where $\disc(\tilde{\cvr}, \cvr, \iota)=2$,
  \item  $u_1$ identifiers $\iota$ where $\disc(\tilde{\cvr}, \cvr, \iota)=-1$,
  \item  $u_2$ identifiers $\iota$ where $\disc(\tilde{\cvr}, \cvr, \iota)=-2$,
    \end{itemize}
  \item[Deletions] There are $d$ identifiers $\iota$ appearing in $\cvr$ that do not appear in $\tilde{\cvr}$.  
  \item[Additions] There are $a$ identifiers $\iota$ appearing in $\tilde{\cvr}$ that do not appear in $\cvr$ or on any ballot.
  \end{description}


\label{def:reinterpreted CVR}
 \end{definition}
 \begin{definition}[Tabulation of CVR]
   Let $\cvr$ be a global CVR for a ballot family $\mathbf{B} = (\mathbf{B}_1, \ldots, \mathbf{B}_k)$. The tabulation
   of $\cvr$ is
   \[
     \Tab(\cvr) = ((\size^{\cvr_1}, \winner^{\cvr_1},
     \loser^{\cvr_1}), ..., (\size^{\cvr_k}, \winner^{\cvr_k},
     \loser^{\cvr_k}))\,.
   \]
   A tabulation $T$ is \emph{consistent} with a
   global CVR $\cvr$ if $T= \Tab(\cvr)$.
 \end{definition}
 
\noindent
Our first claim bounds the (probability distribution of) discrepancy
when the tabulation and CVR are consistent but are inconsistent with
the physical ballots.

\begin{claim} \label{clm:correctness} Let $(o_1, u_1, o_2, u_2, a, d)$
  be natural numbers, let $\mathbf{B} = (\mathbf{B}_1,\ldots,\mathbf{B}_k)$ be a
  ballot family with canonical CVR $\cvr^*$, and let
  $\tilde{\cvr} = (\tilde{\cvr}_1,..., \tilde{\cvr}_k)$ be a
  $(o_1, u_1, o_2, u_2, a, d)$-distortion of $\cvr^*$.
 For a single iteration of
  $\Auditor[\cvrtransform_{\mathrm{Force}}]$
  interacting with $\mathcal{H}(\mathbf{B}, \tilde{\cvr}, \Tab(\tilde{\cvr})))$,
  \begingroup
  \allowdisplaybreaks
\begin{align*}
\frac{o_2-2a-d}{\size^\act} \le \Pr[\disc^\mathcal{H} =2] &\le \frac{o_2+a+2d}{\size^\act},\\
\frac{{o_1-3a-2d}}{{\size^\act}}\le \Pr[\disc^\mathcal{H} =1] &\le \frac{o_1+2a+3d}{\size^\act},\\
\frac{{u_1-3a-2d}}{{\size^\act}}\le \Pr[\disc^\mathcal{H} =-1] &\le \frac{u_1+2a+2d}{\size^\act},\\
\frac{u_2-2a-d}{\size^\act}\le \Pr[\disc^\mathcal{H} =-2] &\le \frac{u_2+a+d}{\size^\act}.
\end{align*}
\endgroup
Furthermore, for $e = o_1+o_2+u_1+u_2$ we have
\begin{align*}
 \frac{1-e-(3a+3d)}{\size^\act}&\le \Pr[\disc^\mathcal{H} =0]\le\frac{1-e+(3a+3d)}{\size^\act}.
\end{align*}
\end{claim}

\begin{proof}
Consider some fixed batch $\batchnum$.  In the absence of additions and delections, overstatement and understatement errors are immediate.  We now consider two cases where the size of the batch is too large and when it is too small.

Let $\size^\cvr_\batchnum > \size^\act_\batchnum$.  Then $\size^\cvr_\batchnum- \size^\act_\batchnum$ rows will be deleted from the $\cvr$. These deleted rows could correspond to any possible discrepancy value.    Note other rows will be adjusted to deal with the discrepancy of the deleted rows.  At most one vote for a winner can be added to a single row and at most one vote for a loser can be added to a single row.  If these are added the same row they do not change the discrepancy.  Otherwise, they increase the discrepancy of one row and decrease the discrepancy of another row.  Thus, to compensate for the removal of a row $2$ instances of a discrepancy of $-1, 0, 1$ can be removed and $2$ added. Compensation can remove two instances of $2,-2$ discrepancy and create at most $1$ row of discrepancy $2, -2$ since a discrepancy of $2,-2$ can never be achieved by subtracting or increasing discrepancy respectively.  This yields the bounds for $a$ in Claim~\ref{clm:correctness}.

Now consider the case when
$\size^\cvr_\batchnum < \size^\act_\batchnum$; then
$\size^\act_\batchnum-\size^\cvr_\batchnum$ rows will be added to the
CVR with identifier $\perp_i$.  Note that the votes on this row can be any value but there will be no matching ballot leading to a discrepancy value of $0, 1$ or $2$. To keep the CVR consistent with the CVR at most $2$ records can have their totals adjusted as with additions. As before, only a single row can be created with a discrepancy of $2,-2$ per deletion.
\end{proof}

Recall that $\cvrtransform_{\mathrm{Force}}$ forces the CVR to be
consistent with the tabulation; thus the transformed CVR has the same
discrepancy as the tabulation with the actual ballots. Ideally, the
observed random variable $\disc$, arising from $\tilde{\cvr}$ under
$\cvrtransform_{\mathrm{Force}}$, would be identical to that arising
from the original CVR $\tilde{\cvr}$.  In the case of only
overstatement and understatement errors this is achieved.

However, this is not achieved in the case of additions and deletions.  Recall that the tabulation and $\tilde{\cvr}$ are consistent. The corrections that happen in $\cvrtransform_{\mathrm{Force}}$ are size corrections due to additions and deletions.  Ideally, $\cvrtransform_{\mathrm{Force}}$ would respond to a deletion by ``adding back'' the deleted row but it has no information about the votes or identifier on the deleted ballot.  Furthermore, any row that is added back may require other rows of the $\tilde{\cvr}$ to be adjusted for consistency with the tabulation.

Similarly, $\cvrtransform_{\mathrm{Force}}$ would ideally respond to
addition by deleting the added row but in general it cannot identify
the added row.  The row it chooses to delete can then yield changes
to the discrepancy distribution as indicated above. Thus, the response to additions can increase or decrease the mean of $\disc$ depending on where they are located.  The response to deletions can never cause a negative discrepancy value because the added row's identifier does not appear on any ballot.

We now consider the case where errors are introduced between the
tabulation and the CVR.  In this setting we assume that the tabulation
has arbitrary disagreements with the canonical CVR so that the effect
of $\cvrtransform_{\mathrm{Force}}$ is to ensure that the CVR for
$\batchnum$ has the same discrepancy as the tabulation.  This means
that the expectation of observed discrepancy will have the same mean
but $\cvrtransform_{\mathrm{Force}}$ can increase the probability that
the observed discrepancy is nonzero, increasing the variance. That is,
errors reduce the chance that the observed discrepancy will be $0$.
In both Claims~\ref{clm:correctness} and \ref{clm:correctness 2} the
actual distribution of discrepancy depends on the distribution of
errors between batches.

\begin{claim} \label{clm:correctness 2} Let
  $(o_1', u_1', o_2', u_2', a', d')$ be natural numbers and let
  $\mathbf{B} = (\mathbf{B}_1,\ldots,\mathbf{B}_k)$ be a ballot family.  Let $T$ be a tabulation for
  $\mathbf{B}$ and let $\cvr_T$ be a uniquely labeled global CVR that
  is consistent with $T$ (so that $T = \Tab(\cvr_T)$). Define
  $d_{-2}, d_{-1}, d_0, d_{1}, d_{2}$ so that for a
  single iteration of $\Auditor[\cvrtransform_{\mathrm{Force}}]$
  interacting with
  $\mathcal{H}(\mathbf{B}, \cvr_T, T)),$
  
  \[
    \forall i, d_i = \Pr[\disc^\mathcal{H} =i] \text{ and }d_e := \sum_i i \cdot d_i.\]
Let $\tilde{\cvr} = (\tilde{\cvr}_1,..., \tilde{\cvr}_k)$ be a $(o_1', u_1', o_2', u_2', a', d')$-distortion of $\cvr_T$.
For a  single iteration of
$\Auditor[\cvrtransform_{\mathrm{Force}}]$
interacting with $\mathcal{H}((\mathbf{B}_1,\ldots,\mathbf{B}_k), \tilde{\cvr}, T)$ one has that 
\begin{align*}
  \Pr[\disc^\mathcal{H} =2] &\in d_2 \pm  \frac{o_2'+o_1'+2u_2'+u_1'+2a'+3d'}{\size^\act},\\
  \Pr[\disc^\mathcal{H}=1] &\in   d_1 \pm \frac{2o_2'+ 2o_1'+2u_2'+2u_1'+2a'+3d'}{\size^\act},\\
   \Pr[\disc^\mathcal{H}=0] &\in d_0 \pm \frac{2o_2'+ 2o_1'+2u_2'+2u_1'+3a'+3d'}{\size^\act},\\  
 \Pr[\disc^\mathcal{H} =-1] &\in   d_{-1} \pm \frac{2o_2' + 2o_1'+2u_2'+2u_1'+2a'+3d'}{\size^\act},\\
 \Pr[\disc^\mathcal{H} =-2] &\in d_{-2} \pm \frac{2o_2'+ o_1'+u_2'+u_1'+2a'+3d'}{\size^\act},
\end{align*}
and $\Exp[\disc^\mathcal{H}]=d_e$.
\end{claim}
\noindent

\begin{proof}
Consider some fixed batch $\batchnum$. 
For a batch with an addition, some row will be deleted which can have an arbitrary discrepancy value.  As in the proof of Claim~\ref{clm:correctness} in the worst case to compensate for the vote totals on the deleted row, one row will have $\winner_r - \loser_r$ increased and another row will have $\winner_r - \loser_r$ decreased.

We now consider deletions. A row may be added which begins with
discrepancy $0$.  The deleted row had an arbitrary discrepancy. When
new rows are added to compensate for the deleted rows the discrepancy
of the $\tilde{\cvr}$ must be adjusted to match the tabulation.
For each deletion, the newly added row can have any vote pattern.  As
before,
 the created row could have a vote pattern different from the ballot that was deleted.  This leads to other ballots having their vote totals adjusted to ensure the total discrepancy between $\tilde{\cvr}_\batchnum$ and the tabulation is $0$.  At most two ballots have to be adjusted to compensate for this created row.  These adjustments can create any discrepancy.

Now consider an $o_2$ error.  This means there is some row $(\iota, 0,1)$ moved
to $(\iota, 1,0)$ in the CVR.  As such in the worst case the checks in
Step~\ref{step:winner consistent} and \ref{step:loser consistent} will
not pass in Figure~\ref{fig:adaptive-auditor} (this would not be the case
if $u_1$ or $u_2$ errors occur in the same batch).  Namely,
$\winner^\cvr> \winner^\tab$ and $\loser^\cvr< \loser^\tab$.  To
compensate for this the procedure in Figure~\ref{fig:transforms}
will change some $\winner$ vote from $1$ to $0$ and some $\loser$ vote
from $0$ to $1$.  If both of these changes happen on the vote with the
$o_2$ error then no problem occurs.  If it happens on two separate this decreases the discrepancy of two rows.
Analysis for the other
cases proceeds in a similar fashion.
\end{proof}


Since the CVR is forced to have the same discrepancy as tabulation, after applying $\cvrtransform_{\mathrm{Force}}$ the produced CVR has the same discrepancy as the tabulation.  But $\cvrtransform_{\mathrm{Force}}$ could increase the probability that discrepancy is nonzero. There are statistical tests that only depend on the expected value of $\obsdisc$.  However, $\Risk$, and thus $\Stop$, of Kaplan-Markov (and many other statistical tests) depends on the entire distribution of $\disc$ (not just its expectation), so these errors do affect stopping time. 


\subsection{Composing the two error models}
\label{ssec:completeness discussion}
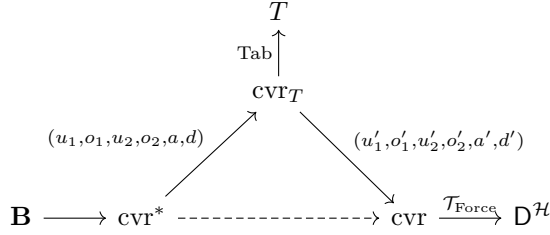
\begin{figure}
  \centering
  \begin{tikzcd}
    && T & \\
    && \cvr_T \arrow[rightarrow,"\Tab"]{u}  \arrow[rightarrow]{dr}{(u_1',o_1',u_2',o_2',a',d')} &&\\[3ex]
    \mathbf{B} \arrow{r} &  \cvr^* \arrow[ur]{dd}{(u_1,o_1,u_2,o_2,a,d)} \arrow[rightarrow,dashed]{rr} && \cvr \arrow["\cvrtransform_{\mathrm{Force}}"]{r}  & \disc^\mathcal{H}
  \end{tikzcd}
 
  \caption{Claim~\ref{clm:correctness} bounds the probability of each discrepancy value for the case when errors are introduced from canonical CVR and tabulation.  Claim~\ref{clm:correctness 2} bounds the probability of each discrepancy value for the case when (additional) errors are introduced from tabulation to the produced batch CVRs.}
     \label{fig:composed correctness bounds}
     \vspace{-.2in}
\end{figure}

Figure~\ref{fig:composed correctness bounds} describes a comprehensive
error model where errors are first added from the canonical CVR and
the tabulation and then further errors are added to the CVRs provided
to the honest adversary.  The bounds obtained by composing
Claims~\ref{clm:correctness} and \ref{clm:correctness 2} are below:
\begin{align*}
  \Pr[\disc^\mathcal{H} =2] &\in o_2 \pm  \frac{2a+ 2d+o_2'+o_1'+2u_2'+u_1'+2a'+3d'}{\size^\act},\\
  \Pr[\disc^\mathcal{H}=1] &\in   o_1 \pm \frac{3a+3d+ 2o_2'+ 2o_1'+2u_2'+2u_1'+2a'+3d'}{\size^\act},\\
 \Pr[\disc^\mathcal{H} =-1] &\in   u_{1} \pm \frac{3a+2d+2o_2' + 2o_1'+2u_2'+2u_1'+2a'+3d'}{\size^\act},\\
 \Pr[\disc^\mathcal{H} =-2] &\in u_{2} \pm \frac{2a+d+2o_2'+ o_1'+u_2'+u_1'+2a'+3d'}{\size^\act},\\
 \Pr[\disc^\mathcal{H}=0] &\ge 1
   - \frac{(o_2+o_1+u_1+u_2)+2(o_2'+ o_1'+u_2'+u_1')}{\size^\act}-\frac{3(a+d+a'+d')}{\size^\act}.
   \end{align*}

Ideally one would show error bounds for an arbitrary combination of ballots, tabulation, and global CVR. Our bounds assume errors are added to global CVRs in two stages first to tabulation, and then to CVRs returned in the audit.  We found global CVRs for each stage to be the most natural way to track differences.  This leads to final bounds that assume a particular distorted CVR used to produce the tabulation that is not seen by any party.

\section{Adaptive Group Comparison Audits}
\label{sec:batch comparison}
We described a methodology to perform ballot
comparison audits without the need to generate a global CVR for the
entire election.  As described in the introduction, no such CVR is
necessary if one wishes to perform a batch comparison audit in
settings where tabulated totals are available for the relevant
batches.
In this section, we show that a hybrid of these techniques is possible
that permits tabulated batches to be broken into smaller untabulated
collections that we call \emph{groups}; these groups of ballots are
then treated analogously to individual ballots in the adaptive audit. In
particular, an audit can hand-count appropriately selected groups and
compare these against an adaptively generated ``group CVR'' that declares
totals for each group. This yields a trade-off between the size of the
groups (and hence the effort involved in hand counting them) and the
number of groups. Ballots do not need to be given identifiers in
this procedure, though groups must be identifiable.

\paragraph{Batch comparison audits}
We begin by reviewing conventional batch comparison audits,
the third major family of risk-limiting audits used in practice.  We borrow notation from Definition~\ref{def:election}.  For an election $E$, a batch comparison audit consists of multiple iterations of the following experiment:
\begin{enumerate}
\item A batch is selected with probability proportional to size.
  \item A full hand count is conducted for the batch.  
  \item The observed discrepancy between the tabulated totals and the hand count is computed.
\end{enumerate}



The envisioned hybrid audit procedure is as follows:
\begin{enumerate}
\item A batch is selected with probability proportional to size.
  \item The batch is separated into $\nu$ groups and an untrusted ``group CVR'' is generated.  
    \label{step:CVR batch}   This CVR reports the size, vote total for $\winner$, and vote total for $\loser$ for each group in the selected batch.  Thus the CVR consists of $\nu$ triples $(\size_{\batchnum,\groupnum}, \winner_{\batchnum, \groupnum}, \loser_{\batchnum, \groupnum})$, one for each value of $\groupnum \in [\nu]$.
  \item A group $\groupnum$ is selected with probability proportional to its purported size, $\size_{\batchnum, \groupnum}$.
  \item A full hand count is conducted for group $\groupnum$.  Let
    $\size^{\act}_{\batchnum, \groupnum}$,
    $\winner^\act_{\batchnum, \groupnum}$, and
    $\loser^{\act}_{\batchnum, \groupnum}$ denote the size and
    relevant totals.
    \item The observed discrepancy is 
    \[
    \obsdisc := \frac{((\winner_{\batchnum, \groupnum}^\cvr -
    \loser^\cvr_{\batchnum, \groupnum}) - (\winner^\act_{\batchnum, \groupnum} - \loser^\act_{\batchnum, \groupnum}))}{\size_{\batchnum, \groupnum}}.
    \]
  \end{enumerate}
  
 Such a procedure may be preferable to batch comparison audits as one
  effectively identifies groups of ballots rather than individual
  ballots. Additionally, as the number of groups is typically much
  smaller than the number of ballots, it may be easier to identify and
  locate a particular group of ballots rather than identify an
  individual ballot.
  Of course, each comparison step in such an audit requires hand
  counting an entire group. 

  The sizes of groups declared in the group CVR is not assumed to be
  correct.
  Note, however that the notion of batch and the assumptions
  pertaining to batches---in particular that a correct manifest is
  supplied to the auditor---are common in the two approaches.

\subsection{Adapting the Formalism}
We now introduce a second \emph{Auditor--Adversary} game for adaptive
group comparison audits.  The relevant notions of election, vote
totals, and ballot manifest are identical to those of
Section~\ref{sec:adversary model}, though ballot identifiers are
irrelevant for this approach. (Rather than formally redefine the
notion of ballot collection to remove identifiers, we leave the notion
unchanged and remark that they are unused.) The meaning of a CVR is
adapted as indicated above so that it declares sizes and vote totals
for groups in a batch (but contains no information about individual
ballots). Figure~\ref{fig:auditing game group} describes the adaptive
batch RLA game between the auditor and adversary.

\begin{definition}[Group Cast-Vote Record (CVR) syntax.] Let
  $E = (\mathbf{B}, T)$ be an election. A
  \emph{Group Cast-Vote Record Table (CVR)} for batch $\batchnum$ of $\numgroups$ groups is a sequence of tuples
\begin{equation*}
  \cvr_\batchnum = ((\size_1^\cvr, \winner_1^\cvr, \loser_1^\cvr),
  \ldots, (\size_\numgroups^\cvr, \winner_\numgroups^\cvr,
  \loser_\numgroups^\cvr))\,, \label{eq:group cvr format}
\end{equation*}
where each coordinate is a natural number.  We borrow
general notation from Definition~\ref{def:cvr format}. We say that a
CVR is \emph{well-formed} if
$\forall \groupnum \in [\numgroups]$ it holds that
$\max(\winner_\groupnum^\cvr,\loser_\groupnum^\cvr) \le \size_\groupnum^\cvr$.
\label{def:group cvr format}
\end{definition}

\begin{figure}[t]
  \begin{framed}
Auditor ($\mathcal{C}$)--Adversary ($\mathcal{A}$) game for
    election $E = (\mathbf{B}, T) $
    \begin{enumerate}[noitemsep]
    \item \textbf{Setup}.
    \begin{enumerate}[noitemsep]
    \item\textbf{Ballot and tabulation delivery (to $\mathcal{A}$).} The
      physical ballots $\mathbf{B}$ and the tabulation $T$ are given to the
      adversary $\mathcal{A}$.
    \item\textbf{Ballot manifest and tabulation delivery
        (to $\mathcal{C}$).}  The ballot manifest
      $\size_E = (\size_1^\act, \ldots, \size_k^\act)$ and the tabulation $T$ are
      given to the auditor $\mathcal{C}$.
    \end{enumerate}
  \item \textbf{Audit}. $\mathcal{C}$ repeatedly makes one of the following two requests of $\mathcal{A}$, or chooses to conclude the
    audit:
    \begin{itemize}
      \itemsep0em
      \itemindent-1em
    \item \textbf{Group CVR request}. For some $\batchnum$,
      $\mathcal{C}$ requests a CVR for batch $\batchnum$.  If the
      batch is not yet partitioned, $\mathcal{A}$ selects a natural
      number $\numgroups \geq 1$ and indelibly assigns each ballot
      $\mathbf{b}\in \mathbf{B}_\batchnum$ to a group
      $\groupnum \in [\numgroups]$.  Denote the partition of groups
      that arise from this assignment
      $\mathbf{B}_{\batchnum, 1},\ldots, \mathbf{B}_{\batchnum,
        \numgroups}$.      $\mathcal{A}$ responds with a group CVR denoted $\CVR_{\batchnum}$.
    \item \textbf{Group request} For some batch $\batchnum$ that has been partitioned into $\numgroups$ groups by $\mathcal{A}$, the auditor  $\mathcal{C}$ requests the physical ballots for a particular group $\groupnum \in [\numgroups]$.  $\mathcal{A}$ responds with $\mathbf{B}_{\batchnum, \groupnum}^* \subseteq \mathbf{B}_{\batchnum, \groupnum}$.
    \end{itemize}
  \item \textbf{Conclusion}. $\mathcal{C}$ returns one of the two values:
$\consistent$ or
$\inconclusive.$
\end{enumerate}
\end{framed}
\caption{The $\RLA_{\mathbf{Group}, \mathcal{C},\mathcal{A}}(E)$ auditing game.}
\vspace{-.1in}
\label{fig:auditing game group}
\end{figure}

At certain points in the security game, the adversary must partition
the ballots from a batch into groups. Once the batch is 
partitioned, this decision is immutable; the adversary may not change the partitioning later. Furthermore, when a group is requested by the
auditor, we require that the adversary responds with a subset of the
selected group. (Equivalently, one may think of the ballots as being
indelibly assigned to groups in such a way that the auditor can
determine the group to which a ballot is assigned and so detect any
situation where the adversary might attempt to include in his response
a ballot from another group.)
Soundness for the above game is as in Definition~\ref{def:risk}: an auditor is $\alpha$-\emph{risk limiting} if for any invalid election $E$ and any
    adversary $\mathcal{A}$,   
    \[
      \Pr_{\mathcal{C}}[\RLA_{\mathbf{Group}, \mathcal{C},\mathcal{A}}(E) = \consistent] \leq \risk\,.
    \]

\subsection{The Auditor}
\label{sec:adaptive group auditor}
We now present an auditor for the adaptive group setting in Figure~\ref{fig:adaptive group auditor} (which adapts Figure~\ref{fig:adaptive-auditor}).  As before, to argue soundness, we consider an identity CVR transform function $\cvrtransform_{\mathrm{Id}}$.

\begin{figure}[th!]
  \begin{framed}
    \underline{Auditor $\mathcal{C}[\cvrtransform,(\Stop,\Criterion)]$ for an
    election $E$}
    \begin{enumerate}[noitemsep]
 \item Receive ballot manifest and tabulation:
       \begin{align*}
        \size_E^\act = (\size_1^\act, \ldots, \size_k^\act);\qquad
        T = (\size^\tab_1;\winner_1^\tab,\loser_1^\tab), \ldots, (\size^\tab_k;
            \winner_k^\tab, \loser_k^\tab))\,.
      \end{align*}
    \item For $\batchnum=1$ to $k$: {} 
      \begin{minipage}[t]{5cm} \vspace{-.73\baselineskip}
        \begin{enumerate}[leftmargin=4.5mm,noitemsep]
        \item $\size_\batchnum^\tab := \size_\batchnum^\act$;
        \item $\winner_\batchnum^\tab := \min(\winner_\batchnum^\tab,\size_\batchnum^\act)$;
        \item $\loser_\batchnum^\tab := \min(\loser_\batchnum^\tab,\size_\batchnum^\act)$.
        \end{enumerate}\end{minipage}
           \item $\displaystyle \begin{aligned}[t]
        \text{Let} \quad S^\act, S^\tab  &:= \sum_{\batchnum=1}^k \size^\tab_\batchnum= \sum_{\batchnum=1}^k \size^\act_\batchnum .\\
        \mu &:= \frac{\sum_{\beta=1}^k(\winner^\tab_\beta - \loser^\tab_\beta)}{\size^\act}\,.
      \end{aligned}$
      \item If $\mu\le 0$ return $\mathtt{Inconclusive}$.

  \item Initialize $\iter=0$.  
    \item Repeat until $\Stopmu(\disc_1, \ldots, \disc_\iter) = 1$:
      \begin{enumerate}[leftmargin=2mm,noitemsep]
      \item Increment $\iter := \iter + 1$.
      \item Perform $\disc_{\iter} := \mathtt{BasicExperiment}$
      \end{enumerate}
  \item If $\Criterion_\mu(\disc_1,\ldots, \disc_\iter)=1$ return $\mathtt{Consistent}$\\ else return $\mathtt{Inconclusive}$.
\end{enumerate}

\underline{$\mathtt{BasicExperiment}$}:
    \begin{enumerate}[noitemsep]
    \item Select batch $\batchnum$ with probability $\size^\tab_\batchnum/ \size^\tab$.
    \item Request CVR for batch $\batchnum$.  Response denoted $\cvr_\batchnum$. \label{step:group start basic}
    \item Apply the transform: $\cvr_\beta := \cvrtransform(\size_E^\act, T, \cvr_{\batchnum})$.
    \item Pick $\groupnum$ with probability $\size_{\batchnum,\groupnum}/\size^\tab_\batchnum$. 
    \item If $\mathtt{CheckConsistent}(\size_E^\act, T, \cvr_\batchnum) = \mathtt{Error}$, Return $2$.
     \item Ask adversary for ballot group $\groupnum$ from batch $\batchnum$.  
     \item Let $\mathbf{B}_{\batchnum, \groupnum}$ denote the returned ballots.  \label{step:group request ballots}
     \item If $|\mathbf{B}_{\batchnum, \groupnum}|\neq \size_{\batchnum, \groupnum}$, return $2$.  
     \item Let $\winner^\act, \loser^\act \in \N$ denote the vote totals
       of the ballots returned by the adversary.
    \item Return 
    $((\winner_\groupnum^\cvr -
    \loser^\cvr_\groupnum) - (\winner^\act - \loser^\act))/\size_{\batchnum, \groupnum}.$
    \label{step:check consistent group}
    
    \label{step:group stop basic}

    \end{enumerate}

\underline{$\mathtt{CheckConsistent}(\size_E^\act, T, \cvr_\batchnum)$}:
\begin{enumerate}[noitemsep]
\item If $\cvr_\batchnum$ is not well formed (Def.~\ref{def:group cvr format}) return $\mathtt{Error}$.
\item If $\size^\cvr_\batchnum, \size^\act_\batchnum, \size^\tab_\batchnum, \sum_\groupnum \size^{\cvr}_{\batchnum, \groupnum}$ are not all equal, return $\mathtt{Error}$.
\item If $\sum_\groupnum \winner_{\batchnum,\groupnum}^\cvr \neq \winner^\tab_\batchnum$ or $\sum_{\groupnum} \loser^\cvr_{\batchnum,\groupnum} \neq \loser^\tab_\batchnum$, return $\mathtt{Error}$.
\item Return $\mathtt{OK}$.
\end{enumerate}

\underline{$\cvrtransform_{\mathrm{Id}}(\size_E^\act, T, \cvr_\batchnum)$}:
\begin{enumerate}[noitemsep]
\item Return $\cvr_\batchnum$.
\end{enumerate}
\end{framed}
\caption{The auditor $\mathcal{C}_{\cvrtransform, (\Stop, \Criterion)}$ for adaptive group comparison.}
\label{fig:adaptive group auditor}
\end{figure}

Next we show that $\mathtt{BasicExperiment}$ yields a
$\disc/|\mathtt{B}|$-dominating random variable $\obsdisc$. Similarly
to the treatment of Claim~\ref{clm:local} for ballot comparison
audits, we begin by focusing on the conditional distribution arising
from fixing a particular batch $\batchnum$ (in the first step of
$\mathtt{BasicExperiment}$). We let
$\mathtt{BasicExperiment}_\batchnum$ refer to this experiment and let
$\disc_\beta^\Adversary$ denote the random variable that arises at the
conclusion of the experiment. As in the analysis of
Claim~\ref{clm:local}, observe that $\groupnum$ is independent of the
partitioning and CVR generated by the adversary. The analysis of the
full experiment $\mathtt{BasicExperiment}$ then follows by linearity
of expectation (Claim~\ref{clm:basic group}). We implicitly work in the context of an arbitrary, but fixed, election $E$ with the constraints and assumptions arising from the portion of the audit preceding the batch and group sampling iterations.

\begin{claim}
\label{clm:local group}
Consider $\mathtt{BasicExperiment}_\batchnum$ in the context of an
election $E = (\mathbf{B}, T)$. Then
  \[
    \Exp[\obsdisc_\batchnum] \geq \disc_\batchnum/\size_\batchnum\,.
  \]
\end{claim}

\begin{proof}
  Let $\mathbf{B}_{1},..., \mathbf{B}_{\nu}$ be the partition of
  ballots created by the adversary for batch $\batchnum$ and let $\cvr$ be the CVR returned
  by the adversary.  We prove the claim for an arbitrary, fixed choice
  of $\cvr$ and $(\mathbf{B}_{\groupnum})_{\groupnum=1}^\nu$; the
  claim then holds for any distribution over these
  values. Recall that
  $\sum_{\groupnum=1}^\numgroups |\mathbf{B}_{\groupnum}| =
  \size_{\batchnum}$.  Note that if
  $\mathtt{CheckConsistent} = \mathtt{Error}$ then
  $\obsdisc_\batchnum=2$. The claim is clearly true in this case since
  $\disc_\batchnum/\size_\batchnum \leq 2$ by definition. We work with
  the assumption $\mathtt{CheckConsistent} = \mathtt{OK}$, and hence $\sum_{\groupnum=1}^{\numgroups} \size_{\batchnum,\groupnum} = \sum_{\groupnum=1}^\numgroups |\mathbf{B}_{\groupnum}| =
  \size_{\batchnum}$, for the
  remainder of the proof.

  In general, for a partition $(\mathbf{A}_1, \ldots, \mathbf{A}_\nu)$
  of the ballots in $\mathbf{B}_\batchnum$ and a family of ballot
  subsets $(\mathbf{A}_1^*, \ldots, \mathbf{A}_\nu^*)$ with the
  property that
  $\forall \groupnum, \mathbf{A}_\groupnum^* \subset
  \mathbf{A}_\groupnum$, we let
  $\disc_\batchnum( (\mathbf{A}_{\groupnum})_{\groupnum=1}^\nu;
  (\mathbf{A}_{\groupnum}^*)_{\groupnum=1}^\nu)$ denote the random
  variable arising from the experiment if the adversary initially
  forms the partition given by $\mathbf{A}_\groupnum$, sends $\cvr$ to
  $\Auditor$, and then answers any request for group $\groupnum$ with
  $\mathbf{A}_\groupnum^*$. We let
  $(\mathbf{B}^*_\groupnum)_{\groupnum=1}^\nu$ be the set family
  determined by the adversary $\Adversary$ so that by definition
  $\obsdisc_\beta =
  \disc_\beta((\mathbf{B}_\groupnum)_{\groupnum=1}^\nu;(\mathbf{B}^*_\groupnum)_{\groupnum=1}^\nu)$. The sets $\mathbf{B}_\groupnum^*$ might not cover
  all the ballots in $\mathbf{B}_\batchnum$.

  We now show that there exists a partition of ballots
  $(\mathbf{B}_{\groupnum}^{\min})_{\groupnum=1}^\nu$ with the
  property that
  $\forall \groupnum, |\mathbf{B}_{\groupnum}^{\min}| =
  \size_{\groupnum}$ and, moreover,
  $\obsdisc_\beta \geq
  \disc_\beta((\mathbf{B}^{\min}_\groupnum)_{\groupnum=1}^\nu;(\mathbf{B}_\groupnum^{\min})_{\groupnum=1}^\nu)$
  (with certainty over choice of $\groupnum$). (Note that in this experiment the same set system is used for the initial partition and the answers of the adversary to group requests.)
  To define the partition $(\mathbf{B}^{\min}_\groupnum)_{\groupnum=1}^\nu$:
  \begin{itemize}
  \item We say that a group $\groupnum$ is viable
    $|\mathbf{B}^*_\groupnum| = \size_\groupnum$. In this case, define
    $\mathbf{B}^{\min}_\groupnum = \mathbf{B}_\groupnum^*$. Let $\mathbf{B}_{\text{viable}} = \bigcup_{\groupnum | \groupnum\text{ is viable}} \mathbf{B}_\groupnum^*$.
  \item The sets $\mathbf{B}^{\min}_\groupnum$ for nonviable
    $\groupnum$ are defined to form an arbitrary partition of the
    remaining ballots
    $\mathbf{B}_\batchnum \setminus \mathbf{B}_{\text{viable}}$ with
    the size constraints
    $\forall \,\text{nonviable $\groupnum$}, |\mathbf{B}_\groupnum^{\min}| =
    \size_\groupnum$. Note that this is always possible because $\sum \size_{\beta,\groupnum} = |\mathbf{B}_\batchnum|$.
  \end{itemize}

Any size mismatch (when the subset of ballots returned by the adversary for a request for group $\groupnum$ does not have size $\size_{\batchnum,\groupnum}$) results in a maximal, default
discrepancy of 2. It follows that
\[
  \obsdisc_\beta =
  \disc_\beta((\mathbf{B}_\groupnum)_{\groupnum=1}^\nu;(\mathbf{B}^*_\groupnum)_{\groupnum=1}^\nu) \ge
  \disc_\batchnum( (\mathbf{B}_{\groupnum}^{\min})_{\groupnum=1}^\nu), (\mathbf{B}_{\groupnum}^{\min})_{\groupnum=1}^\nu) \,.
\]
Specifically, note that $\groupnum$ is drawn according to the same
distribution in the two experiments and, for any viable $\groupnum$, these two
random variables take the same value; for any nonviable $\groupnum$ the first
takes the default value of $2$, while the second is
\[
\frac{(\winner^\cvr_{\groupnum} - \loser^\cvr_{\groupnum}) - (\winner_{\groupnum}^\act - \loser_{\groupnum}^\act)}{\size_{\groupnum}} \leq 2\,,
\]
where the actual vote totals here are with respect to $(\mathbf{B}^{\min}_\groupnum)$.
Then one has that
\begingroup
  \allowdisplaybreaks
\begin{align*}
      \Exp\left[\obsdisc_\batchnum\right] &\geq 
         \Exp\left[\obsdisc_\batchnum((\mathbf{B}_{\groupnum}^{\min})_{\groupnum=1}^\nu),(\mathbf{B}_{\groupnum}^{\min})_{\groupnum=1}^\nu)\right]\\  
         &=\sum_{\groupnum=1}^\nu \frac{\size_{\groupnum}}{\size_\batchnum} \left( \frac{(\winner^\cvr_{\groupnum} - \loser^\cvr_{\groupnum}) - (\winner_{\groupnum}^\act - \loser_{\groupnum}^\act)}{\size_{\groupnum}}\right)\\
                  &=\frac{(\winner^\cvr_\batchnum - \loser^\cvr_\batchnum)}{\size_\batchnum} - \sum_{\groupnum=1}^\nu \frac{1}{\size_\batchnum}   (\winner_{\groupnum}^\act - \loser_{\groupnum}^\act)\\
                                    &=\frac{(\winner^\cvr_\batchnum - \loser^\cvr_\batchnum)}{\size_\batchnum} -\sum_{\groupnum=1}^\nu \frac{1}{\size_\batchnum} \left(\sum_{\mathbf{b} \in \mathbf{B}_{\groupnum}^{\min}} (\winner_{\mathbf{b}}^\act - \loser_{\mathbf{b}}^\act)\right)\\
                                                                        &=\frac{1}{\size_\batchnum}\left((\winner^\cvr_\batchnum - \loser^\cvr_\batchnum) - \sum_{\mathbf{b} \in \mathbf{B}_{\batchnum}} (\winner_{\mathbf{b}}^\act - \loser_{\mathbf{b}}^\act)\right)\\
      &= \frac{1}{\size_\batchnum} \left( (\winner^\tab_\batchnum - \loser^\tab_\batchnum) - (\winner_\batchnum^\act - \loser_\batchnum^\act)\right)
      = \frac{\disc_\batchnum}{\size_\batchnum}\,.
\end{align*}
\endgroup
  This completes the proof of Claim~\ref{clm:local group}.
\end{proof}
Showing that this extends to the overall discrepancy follows exactly as in Claim~\ref{clm:one iter}: 

\begin{claim} \label{clm:basic group} The expectation of $\obsdisc$ over a single iteration satisfies
  \begin{align*}
    \Exp[\obsdisc] = \sum_\batchnum \left(\frac{\size^\man_\batchnum}{\size^\man} \cdot
    \Exp[\obsdisc_\batchnum] \right) \geq \sum_\batchnum \left(\frac{\size^\man_\batchnum}{\size^\man} \cdot
    \frac{\disc_\batchnum}{\size_\batchnum^\man}\right) = \frac{\disc}{\size}\,.
\end{align*}
\label{clm:one iter group}
\end{claim}
 Furthermore, one can easily show that CVR transforms do not affect whether the auditor is risk-limiting as in Lemma~\ref{lem:permissive okay}.

\paragraph{Why group sizes don't have to be trusted.} Our techniques for trusting an adversarial declaration of group sizes do not extend to an adversarial declaration of batch sizes which must still be counted or verified by a trustworthy component.  There are two key differences in the group setting:
\begin{enumerate}
\item Group size is only hand-counted if selected, and
\item An iteration is marked with $\disc=2$ on any size mismatch.
\end{enumerate}

In principle in an adaptive ballot comparison audit, one could add these two steps of first-hand counting the entire batch and rejecting if the true size is not equal to the declared size.  However, we expect this to be drastically more work and likely to introduce more errors given the larger size of batches.  One could use this technique for small batches, for example, ballots at a precinct that contain votes for valid write-in candidates are often tabulated separately.

\section{Conclusion}
\label{sec:conclusion}

This article presents a formal model of comparison risk-limiting audits and a new class of risk-limiting audits called adaptive
 comparison audits. The formal model allows us to answer critical procedural questions such as showing that the labeling of ballots need not be trusted.  Adaptive comparison audits provide efficiency improvements as one only produces a CVR for batches selected for audit.

\section*{Acknowledgments}
These results were developed as part of a collaboration with the Office of the CT Secretary of State and, additionally, were supported in part by a grant from that office.

Discussions with Mark Lindeman, Philip B.\ Stark, Lynn Garland, and anonymous reviewers improved the narrative and technical treatment.  A.R. is supported by a research grant from IOG and NSF grant \#1801487; B.F. is supported by NSF Grants \#2232813 and \#2141033 and the Office of Naval Research.

\bibliographystyle{abbrv}
\bibliography{RLA}

\appendix

\section{Calculation of CVR Generation Percentages}
\label{sec:statistical analysis}

In this section, we discuss the reported percentages of CVR generated with the adaptive ballot comparison method.  We use Connecticut and Florida as case studies for three reasons:
\begin{enumerate*}
\item elections are managed by each municipality with no voting equipment that is capable of producing CVRs with identifiers, 
\item they represent different population sizes and number of precincts with Florida having approximately $6000$ precincts and Connecticut having approximately $700$, and 
\item there is a large variance in municipality size.
\end{enumerate*}
Furthermore, Connecticut uses a semi-automated transitive tabulator~\cite{antonyan2013computer} to produce CVRs after the fact for some fraction of municipalities.

Our experimental framework adopts the Kaplan-Markov test presented in Definition~\ref{def:km} with $\gamma = 1.1$ and ``a bit of rounding''~\cite{lindeman2012gentle}. In particular, ballot sample sizes were obtained from Neal McBurnett's tool, \texttt{rlacalc}~\cite{nealcode}, using the following data:
\begin{enumerate*}
\item For Connecticut, the number of ballots used is 1,823,857, which is the number of votes cast in 2020 CT presidential election. 
\item For Florida, the population of ballots is 11,067,456, which is the number of votes cast in the 2020 FL presidential election. 
\end{enumerate*}


The number of precincts and voters for each town is pulled from the Connecticut Secretary of State's website and Florida's precinct-level election results.  Ballots were split among towns by reserving 5\% of votes as absentee and then splitting the remaining 95\% evenly into the number of precincts in that town. This means that for a town the number of batches is always one more than the number of precincts. $100$ simulations are conducted of the following experiment:
\begin{enumerate}
\item Randomly distribute ballots to precincts according to their size.
\item Randomly pick (with replacement) sample size ballots among all ballots.  For all batches with a picked ballot mark the batch as picked
\item Compute the total fraction of ballots in batches that are picked divided by the total number of ballots. 
\end{enumerate}
This last fraction is reported as the fraction of CVR generated.  We report the average value of number of distinct picked batches and fraction of
generated CVR are summarized in Table~\ref{tab:costs}. The full
simulation software is available at
this \href{https://github.com/aeharrison815/Adaptive-RLA-Tools}{Github
  repository}.  The full simulation code can also \begin{enumerate*} \item distribute overstatement and understatement errors, and \item compute risk and stopping time.\end{enumerate*}  However, this functionality was not used to create Table~\ref{tab:costs}.

\section{Auditor and transform without overvotes}

\label{app:exclusive}

In Section~\ref{sec:auditor} we presented an auditor that allows ``overvotes''~\cite{lindeman2012gentle}.  An overvote means that a CVR row or ballot that has marks for both the winner and loser is considered valid. It is also possible for $\cvrtransform_{\mathrm{Force}}$ to create overvotes.  

Here we present an alternative auditor and transform function that does not allow or create overvotes.  The auditor differs from Figure~\ref{fig:adaptive-auditor} in exactly two places:
\begin{enumerate}
\item Step~\ref{step:correct winner} which sets $\winner_\iter^\tab := \min(\winner_\iter^\tab,\size_\iter^\act)$ is moved after Step~\ref{step:correct loser} and replaced with $\winner_\iter^\tab := \min(\winner_\iter^\tab,\size_\iter^\act -\loser_\iter^\tab)$.  This ensures that the sum of $\winner_\iter^\tab+\loser_\iter^\tab\le \size_\iter^\act$.  
\item A check is added to $\mathtt{CheckConsistent}$ as follows: If there exists a row with identifier $\iota$ in $\cvr_\batchnum$ such that $\winner_\iota =1$ and $\loser_\iota =1$ return $\mathtt{Error}$.  This step is added before the step that returns $\mathtt{OK}$.  Let $\mathtt{CheckConsistent}_\overvote$ denote the modified procedure.
\end{enumerate}   The main changes are in the transform function shown in Figure~\ref{fig:transforms overvote} here the transform never creates a row where both winner and loser are $1$.  Differences are highlighted in Blue.

\begin{figure}[t]
  \begin{framed}
\underline{$\cvrtransform_{\overvote, \mathrm{Force}}(\size_E^\act, T, \cvr_\batchnum)$}:
\begin{enumerate}[noitemsep]
\item If $\cvr_\batchnum$ is not properly formed tuple according to Definition~\ref{def:cvr format} output $\mathtt{Error}$.
\item  While there exist two rows $i$ and $j$
  where $i<j$ and both have identifier $\iota$, replace the identifier
  in row $j$ with an unused identifier in $\{\badrow_t\}$.
\item \label{step:sizes consistent overvote} If $\size_\batchnum^\cvr \neq \size_\batchnum^\act$, then 
\begin{enumerate}[noitemsep]
\item While $\size_\batchnum^\cvr< \size_\batchnum^\act$ add a new row
  to $\cvr_\batchnum$ with an unused identifier in $\{\badrow_t\}$ and
  zeroes for all votes.
\item While $\size_\batchnum^\cvr> \size_\batchnum^\act$ remove the last row of $\cvr_\batchnum$.  
\end{enumerate}
\item {\color{blue} Place all rows with  $\iota \in\{\badrow_t\}$ at the end of the CVR.}
\item {\color{blue} For all $\iota$ where  $\winner_{r_\iota} =1, \loser_{r_\iota}=1$ set $\winner_{r_\iota}=0$.}
\item \label{step:winner consistent overvote} If $\winner_\batchnum^\cvr \neq \winner_\batchnum^\tab$.
\begin{enumerate}[noitemsep]
\item While $\winner_\batchnum^\cvr< \winner_\batchnum^\tab$ 
\begin{enumerate}[noitemsep]
\item {\color{blue} While $\loser_\batchnum^\cvr>\loser_\batchnum^\tab$, find the last row $r$ such that $\loser_r=1$ set $\winner_r=1, \loser_r=0$.}
\item {\color{blue} Find the last row $r$ such that $\winner_r = 0, \loser_r=0$ set $\winner_r=1, \loser_r=0$.}
\end{enumerate}
\item While $\winner_\batchnum^\cvr> \winner_\batchnum^\tab$ 
\begin{enumerate}[noitemsep]
\item {\color{blue} While $\loser_\batchnum^\cvr<\loser_\batchnum^\tab$, find the last row $r$ such that $\winner_r = 1$ set $\winner_r=0, \loser_r=1$.}
\item {\color{blue} Find the last row $r$ such that $\winner_r = 0, \loser_r=0$ set $\winner_r=0, \loser_r=1$.}
\end{enumerate}
\end{enumerate}
\item \label{step:loser consistent overvote} If $\loser_\batchnum^\cvr\neq  \loser_\batchnum^\tab$. Set $i := \size_\batchnum^\cvr$.
\begin{enumerate}[noitemsep]
\item While $\loser_\batchnum^\cvr< \loser_\batchnum^\tab$:
{\color{blue} find the last row $r$ such that $\winner_r = 0, \loser_r=0$ set $\winner_r=1, \loser_r=0$.}
\item While $\loser_\batchnum^\cvr > \loser_\batchnum^\tab$: {\color{blue} 
    find the last row $r$ such that $\winner_r = 0, \loser_r=1$ set $\winner_r=0, \loser_r=0$.}
  \vspace{-1ex}
\end{enumerate}
\end{enumerate}
\end{framed}
\caption{CVR transform function that ensures consistency and no overvotes.}
\label{fig:transforms overvote}
\end{figure}

\begin{claim}
\label{clm:overvote correction works}
Figure~\ref{fig:transforms overvote} always completes and outputs a CVR such that $\mathtt{CheckConsistent}_\overvote$ returns $\mathtt{OK}$.
\end{claim}
\begin{proof}
Importantly, after Step~\ref{step:correct tabulation} in the modified Figure~\ref{fig:adaptive-auditor} it is true that for all batches $k$, 
\[\winner_k^\tab + \loser_k^\tab\le \size_k^\act.\]
Furthermore, after Step~\ref{step:sizes consistent overvote} in Figure~\ref{fig:transforms overvote} it is true that $\size_k^\act = \size_k^\tab = \size_k^\cvr$. 
We now show that Steps~\ref{step:winner consistent overvote} and \ref{step:loser consistent overvote} in Figure~\ref{fig:transforms overvote} eventually lead to a CVR consistent with the tabulation without overvotes.  
At each iteration of Step~\ref{step:winner consistent overvote} one of four conditions must be true:
\begin{enumerate}
\item $\winner_k^\cvr = \winner_k^\tab$,
\item $\winner_k^\cvr> \winner_k^\tab$,
\item $\loser_k^\cvr>\loser_k^\tab$, or
\item There is a row in the CVR with identifier $\iota$ such that $\winner_\iota^\cvr = 0, \loser_\iota^\cvr=0$.
\end{enumerate}
To see that the four cases are complete, if the first three cases are not true then $\winner_k^\cvr < \winner_k^\tab, \loser_k^\cvr\leq \loser_k^\tab$.  This means that \[\winner_k^\cvr+\loser_k^\cvr< \winner_k^\tab+\loser_k^\tab\le \size_k^\tab = \size_k^\cvr.\]  That is, there are fewer than $\size_k^\cvr$ $1$s in the CVR and there must be some row with both winner and loser set to $0$. 

In each of the above cases, Step~\ref{step:winner consistent overvote} either finds a row to change or completes.  Furthermore, note that $\winner_k^\cvr$ monotonically approaches $\winner_k^\tab$ so it only requires at most $|\winner_k^\tab - \winner_k^\cvr|$ steps to complete.

For Step~\ref{step:loser consistent overvote} note that in addition to the above properties it now holds that $\winner_k^\cvr = \winner_k^\tab$.  Of course, if $\loser_k^\cvr> \loser_k^\tab$ one can always change a row with $\loser_k^\cvr =1$ and $\winner^\cvr=0$ to be both $0$.   Now suppose that $\loser_k^\cvr < \loser_k^\tab$, then it holds that 
\[\winner_k^\cvr+\loser_k^\cvr= \winner_k^\tab+\loser_k^\cvr < \winner_k^\tab+\loser_k^\tab \le \size_k^\tab = \size_k^\cvr.\] That is, there are fewer $\size_k^\cvr$ $1$s in the CVR and there must be some row with both winner and loser set to $0$. 
This completes the proof of Claim~\ref{clm:overvote correction works}.
\end{proof}

\end{document}